\documentclass[useAMS,usenatbib,astrobib]{mn2e}
\usepackage{graphicx}

\usepackage[utf8]{inputenc}
\usepackage{hyperref}
\usepackage{color}
\definecolor{Blue}{rgb}{0,0.08,0.65}
\definecolor{Red}{rgb}{0.65,0.08,0.05}
\definecolor{Green}{rgb}{0.15,0.45,0.25}

\newcommand{\SED}{spectral energy distribution}

\newcommand{\SSP}{single stellar population}

\begin{document}
\title[On the filamentary environment of galaxies  ]{On the filamentary environment of  galaxies
}

\author[C. Gay, C.\ Pichon, D. Le Borgne, R. Teyssier, T. Sousbie, J. Devriendt
 ]
{  C.\ Gay$^{1}$, C.\ Pichon$^{1,5}$, D.\ Le Borgne$^{1}$, R.\ Teyssier$^{2}$, T. Sousbie$^{1,3}$, J. Devriendt$^{4,5}$\\
$^1$  Institut d'Astrophysique de Paris, UPMC Univ Paris 06, CNRS, UMR 7095, F-75014, Paris, France\\
$^2$ Service d'Astrophysique, IRFU,  CEA-CNRS, L'orme des meurisiers, 91470, Gif sur Yvette, France \\
$^3$ Tokyo University, Physics Dept 7-3-1 Hongo Bunkyo-ku,JP Tokyo, 113-0033, Japan\\
$^4$ Observatoire de Lyon (UMR 5574), 9 avenue Charles Andr\'e, F-69561 Saint Genis Laval, France. \\
$^5$ Department of Physics, Denys Wilkinson Building, Keble Road, Oxford, OX1 3RH, United Kingdom. 
}

\date{Typeset \today; Received / Accepted}

\pagerange{\pageref{firstpage}--\pageref{lastpage}} \pubyear{2009}


\maketitle
\label{firstpage}

\begin{abstract}
The correlation between the large-scale distribution of galaxies and their spectroscopic properties at $z=1.5$ is investigated using the Horizon MareNostrum cosmological run.

We have extracted a large sample of $10^5$ galaxies from this large hydrodynamical simulation featuring standard galaxy formation physics. Spectral synthesis is applied to these single stellar populations to generate spectra and colours for all galaxies. We use the skeleton as a tracer of the cosmic web and study how our galaxy catalogue depends on the distance to the skeleton.
We show that galaxies closer to the skeleton tend to be redder, but that the effect is mostly due to the proximity of large haloes at the nodes of the skeleton, rather than the filaments themselves.

This effects translate into a bimodality in the colour distribution of our sample. The origin of this bimodality is investigated and seems to follow from the ram pressure stripping of satellite galaxies within the more massive clusters of the simulation.

The virtual catalogues (spectroscopical properties of the MareNostrum galaxies at various redshifts)
are available online at {\tt http://www.iap.fr/users/pichon/} {\tt MareNostrum/catalogues}. 

\end{abstract}

\begin{keywords}
  large-scale structure of the Universe
  galaxies: evolution
  methods: N-body simulations
  hydrodynamics
\end{keywords}

\section{Introduction}\label{sec:intro}

\begin{figure}
\includegraphics[width= 1.1\columnwidth]{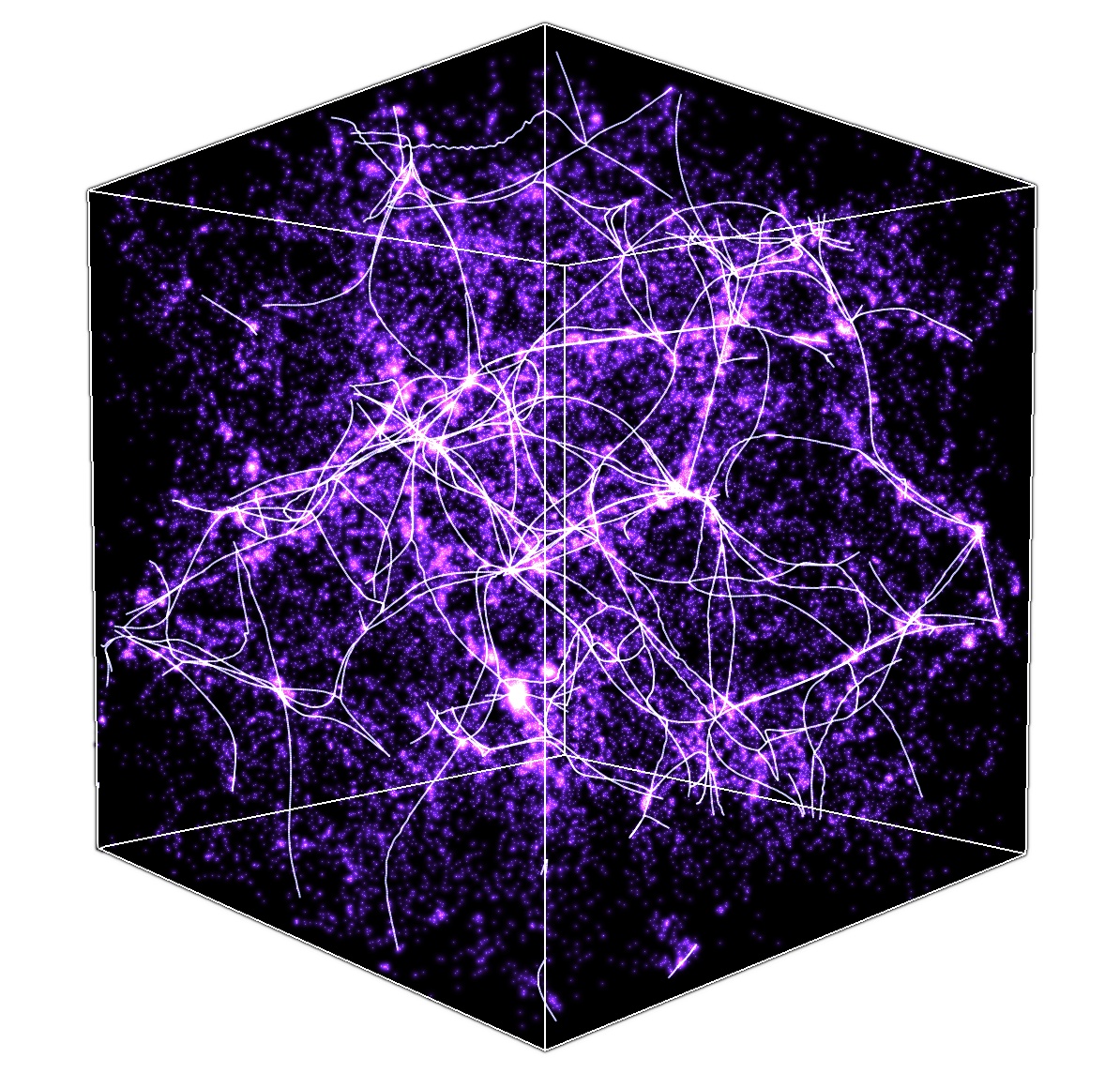}
\caption{A 3D view of the galaxies and the skeleton of the dark matter of  MareNostrum at $z=1.6$.
The box is 50 $h^{-1}$Mpc aside.}
\label{fig:skel} 
\end{figure}

During the past decade, the $\Lambda {\rm {CDM}}$ cosmological model of the Universe has been established as the framework of choice in which to interpret 
how and when observed galaxies acquire their properties. Arguably the most important feature of this framework is to provide us with an explanation as to  
why  many  of  these properties (physical sizes, luminosities) strongly correlate with galaxy mass while others (star formation rates, morphological type) do not seem to.
Unsurprizingly, the all-time favored culprit is the interplay between galaxies and the intergalactic medium (IGM) at large. In other words,  
the large scale environment of galaxies is claimed to play an important role in shaping some of their properties, while the rest of them are thought to depend solely
on small scale (internal) processes. However, having said that, one still has to determine for which of these properties ``Nurture" dominates over ``Nature" and therein 
lies the whole difficulty of the issue.

Indeed, since the early 70s, there has been a plethora of studies devoted to measuring the impact of environment on galaxy properties. 
\cite{DavisGeller1976} first pointed out that early-type galaxies are more strongly clustered than the late types, while \cite{Dressler1980}
and \cite{PostmanGeller1984} demonstrated the existence of a morphology-density relation (MDR). Following in their footsteps, \cite{Balogh1998}, 
\cite{Hashimoto} and more recently \cite{Christlein} and \cite{Poggianti} systematically showed that galaxies living in denser 
environments tend to be redder and have lower star formation rates (SFRs) than their 
more isolated counterparts. One can think of several physical processes associated with different types of environment that could play a role in causing 
such alterations. More specifically, they include, 
by ascending order of environment density: (i) major (wet) mergers, which can turn spiral galaxies into ellipticals (e.g. \cite{toomre}), and 
drive a massive starburst wind 
which quenches future star formation by ejecting the interstellar medium (ISM) out of galaxies (\cite{Mihos,maclow}); (ii) 
active galactic nuclei (AGN) or shock-driven winds (\cite{norm,sprin}; (iii) galaxy ``harassment"
(rapid encounters or fly-bys which dominate over mergers in rich clusters) causing discs to heat and possibly triggerring the build-up of a bulge via the 
formation of a bar (\emph{e.g.} \cite{moore}).
For the latter category, the diffuse gas associated with the galaxies' host dark matter (sub)halo which constitutes the main
 fuel supply for future star formation can also be stripped, thus suppressing later star formation by ``starvation" or ``strangulation" \citep{larson,Bekki}.
Moreover, part of their ISM can also be pulled out of these galaxies, either by tidal forces arising from the gravitational potential of the cluster or by ram pressure 
stripping by the intracluster medium (ICM) \cite{Gunn,Abadi,Chung}.  \\
Although these latter environment dependent processes seem potent enough, recent work carried out by  \cite{Tanaka} and \cite{vdB}  indicate that they
might not be the main mechanisms for quenching star formation activity. This claim is corroborated by the higher redshift results ($ z \!\sim\! 1$) obtained with the DEEP2 
(e.g. \cite{Cooper2006,Cooper2007,Cooper2008,Gerke2007,Coil2008}), (z--)COSMOS \citep{Scoville2007,Cassata2007,Tasca2009} and VVDS \citep{Scodeggio} surveys 
where clusters are more scarce, along with the fact that morphological and spectrophotometric properties of local galaxies are also found to be correlated with 
 their internal properties, such as luminosity, mass or internal velocity (e.g. \cite{Kauffmann2003}). 
Here as well, one can invoke various physical processes to explain such dependences on internal properties. Supernova feedback  
can heat and stir the ISM, possibly ejecting large amounts of gas out of galaxies and it is expected to scale with galaxy 
mass \citep{Larson1974,Dekel1986}. There also exists a growing host of observational evidence that AGN play a key part in quenching star formation 
\citep{Schawinski2006,Schawinski2007,Salim2007} and this AGN feedback should likely impact more massive galaxies since they 
host larger mass black holes \citep{Magorrian,Silk1998}. In light of these investigations, it becomes apparent that 
disentangling nature and nurture is a more complicated process than one would naively have thought to begin with.

Clearly, the fundamental requirement to tackle this issue is to properly characterize the anisotropic  environment of galaxies, both  
in observational samples and theoretical models, spanning as broad a range of environments as possible, from isolated field galaxies 
to groups and rich clusters. The vast majority of the studies in the literature accomplish this task either by counting the number of neighbours that a galaxy has 
within a fixed aperture on the sky or by measuring the distance to the $n^{\rm th}$ nearest galaxy, where $n$ is an integer in the range 3–10. 
Although these indicators are straightforward to obtain, their physical interpretation, let alone their comparison to theoretical models 
are far from being straightforward (\emph{cf.} \cite{Kauffmann2004,Weinmann2006}). Meanwhile, looking at the distribution of observed galaxies in 
modern cosmological surveys, such as the 2dF \citep{2dF} and the SDSS 
\citep{SDSS}, the most striking feature is that they look organised along linear structures linking clusters together (see Figure \ref{fig:skel}). This filamentary 
network, dubbed as the ``Cosmic Web" \citep{bond}, has a dynamical origin and reflects the anisotropic accretion taking place in clusters \citep{skel1}. It therefore 
seems natural to describe the environment of galaxies in terms of their location with respect to these filaments in order to investigate the influence of the Cosmic Web on the 
properties of the galaxies it encompasses.


In this  paper, we carry out such a study at intermediate redshift ($z\sim1.5$), mainly from a theoretical perspective, using a recent diagnostic tool to characterize the 3D environment
called the skeleton \citep{sousb08},
which we combine with the largest  hydrodynamical cosmological simulation performed to date \citep{ocvirk2008,prunet2008, dekelnature}. Run on the MareNostrum computer at the Barcelona
Supercomputer Center using  the RAMSES  code \citep{teyssier02}, this simulation is one of the flagship simulations realized by the  Horizon collaboration ({\tt http://www.projet-horizon.fr}). It includes   a detailed treatment of metal--dependent gas cooling, UV heating, star formation, supernovae 
feedback and metal enrichment.

Specifically, we  will address the question: are the physical conditions \emph{within the filaments} dramatic enough to strongly  influence the properties of the galaxies it 
encompasses?

The  outline  of  this  paper  is  as  follows:  first  we  describe  in Section \ref{sec:method} our
methodology,   in  terms   of  numerical   techniques, estimators  and  statistical
measurements.   The dependence of the spectroscopic properties on the filamentary environment 
is then  discussed in Section \ref{sec:fil}, while Section \ref{sec:bimodality} investigates the observed bimodality and discusses comparison to observations,
and Section \ref{sec:conclusion} wraps up.   Some checks are performed in Appendix A, Appendix B describes the  publicly available catalogues and Appendix C sums up the subgrid physics used.

\section{Methodology}
\label{sec:method}
Let us first describe the MareNostrum simulation, a
cosmological  N  body and  hydrodynamical  simulation of  unprecedented
scale which accounts for most of the physical processes involved in galaxy formation
theory. The procedure
to identify galaxies and  assign spectra and colours to them is also described.
Finally the method used to identify filaments within the simulation is presented.

\subsection{The MareNostrum simulation}
\label{s:MN}
We use the MareNostrum simulation, described in detail in \cite{ocvirk2008}.
To summarise, this simulation uses the AMR code RAMSES \cite{teyssier02} in a periodic box of comoving 50 $h^{-1}$ Mpc, with a $\Lambda$CDM universe ($\Omega_M=0.3$, $\Omega_\Lambda=0.7$, $\Omega_B=0.045$, $H_0=70$~km$\cdot$s$^{-1}\cdot$Mpc$^{-1}$, $\sigma_8=0.9$). The initial grid consists of $1024^3$ dark-matter particles ($m_{\rm part.}  \simeq  8\times 10^6$  $M_{\odot}$) and the same number of cells, which are refined up to five folds when they have more than 8 particles, as long as the the minimum cell size is not under 1 kpc in physical units.
In addition to its large volume, the MareNostrum simulation provides the basic  physical ingredients relevant to galaxy formation. The hydrodynamic physics uses metal-dependent cooling, UV heating (\cite{HM} background model), star formation \citep{RT},
 supernovae feedback and metal enrichment \citep{Dubois08}.
  The ISM, \emph{i.e.} gas with density above 0.1 atom$/{\rm cm}^3$ is modelled with a polytropic equation  \citep{Schaye07} and forms stars consistently with the Kennicutt law with a star formation efficiency of 5\% (see Appendix~\ref{sec:recette}). 
For each snapshot, the dark-matter substructures are detected with the {\tt Adaptahop} algorithm \citep{aubert,tweed} and all stars are associated to a (virtual) galaxy.
The simulation stopped at redshift $z\approx 1.5$, with $5\times10^9$ cells, $1.4 \times10^8$ star particles and around 100,000 galaxies.
The  simulation parameters  ($L_{\rm box}=50$  $h^{-1}$Mpc, $1024^3$  dark
matter particles, and a spatial  resolution close to 1 $h^{-1}$kpc physical) are
optimal  to capture  the most  important spectral properties  of   typical Milky-Way-like galaxies.  The box size allows us
to  have a large  sample  of  about 100,000 $L_{\star}$  galaxies at
redshift  2 and above,  with a  strong statistical  significance (see Appendix~\ref{sec:catalogs} for a detailed description of our sample).  

\begin{figure}
\includegraphics[width= \columnwidth]{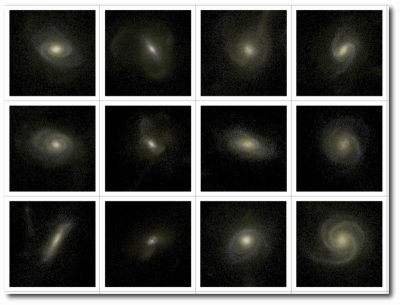}
\caption{a sample of MareNostrum galaxies in the IRAC-8$\mu$m, K, I bands which were used in this investigation.
More such images are available at {\tt http://www.iap.fr/users/pichon/galaxies/}.  }
\label{fig:MNgal}  
\end{figure}


\subsection{Spectral synthesis}
One main interest of the MareNostrum simulation lies in its
ability to  yield realistic virtual observations in a consistent cosmological framework,
which in turn can be compared to real
data.  The outcome of these comparisons should lead to clues about the
physics which drives the evolution of galaxies.

To make such predictions, light needs to be added to the
simulation. This can be done
very naturally by associating spectral synthesis population models to
the star-formation modelling in the MareNostrum
simulation. More precisely, the ISM gas produces a \SSP{} (SSP) per gas cell at the rate described  in section \ref{s:MN}. Each
of these SSPs has the metallicity of the gas which gave birth to
the stars. It is described by a particle in the simulation with a
mass $M\star$ (minimum value $\simeq 10^6 M_\odot$), a metallicity, a
redshift of formation and a position. 
Then, we assign a dust-free \emph{evolving} \SED{} (SED) to each of these
stellar particles with the {\sc PEGASE.2} \citep{PEG,PEG2} population synthesis
code. A Salpeter IMF is used for the spectral modelling, consistently
with the SN feedback.

Finally, the light content of a galaxy at redshift $z$ is the sum of
the SEDs produced by its ``stellar" (i.e. SSP) particles which depend
on their age and metallicity:
\begin{equation}
  F_\lambda (\lambda, z) = \sum_{i=0}^n M^\star_i \!\times\!
  F^\mathrm{1\,M_\odot \mathrm{SSP}}_{\lambda} \left(\lambda,t(z)-t(z^\mathrm{for}_i), Z(z^\mathrm{for}_i)\right)\!,
\end{equation}
where $\lambda$ is the wavelength, $n$ is the number of stellar particles inside the galaxy,
$M^\star_i$ is the total stellar mass of the $i^{th}$ SSP, $t(z)$ is
the Hubble time at redshift $z$ and $z^\mathrm{for}_i$ is the redshift
of formation of the $i^{th}$ SSP, and $Z$ is the metallicity of the
gas.
Here, in contrast to the usual approximation of instantaneous mixing of
chemical elements in a galaxy adopted by many models of galaxy
formation (such as semi-analytic models, e.g. \cite{hatton}), the simulation makes it possible to account for spatial
variations of the metallicity within a galaxy.
Recall that in this paper a galaxy is {\sl defined} to be a set of  at least  10 stars 
which are embedded within a given dark halo subclump. 
\cite{Rimes} presents an alternative definition of a galaxy within the MareNostrum simulation based on 
a threshold in the baryon (gas+star) density; it was checked that 
both definitions yield very similar luminosity functions at various redshifts.
Appendix~\ref{s:modeldamien} shows that the redshift evolution of the corresponding colours seems consistent.
%

We choose not to include reddening by dust in the predicted colours for
the galaxies for two reasons. The first one is that our knowledge of
dust properties and spatial distribution with respect to the gas is
still  somewhat uncertain. The accurate modelling of dust attenuation is a
complex issue and we do not want to enter into debates regarding this 
modelling. 
The second reason is more technical: ray-tracing photons involves
potentially multiple scattering, and such an approach is technically difficult to implement in a very large simulation like
MareNostrum  (see \cite{Devriendt2009} and Appendix  \ref{sec:dust-ok} for an alternative approach). 
In the following, one must keep in mind that the effect of dust on
galaxies' light is not accounted for
in our work. This restriction should  not impact our qualitative findings in terms of  the  influence of filaments,
 provided dust follows light, which seems a good first order approximation for this simulation (\cite{Rimes} and Appendix~\ref{sec:dust-ok}).
 
We do include, however, IGM absorption on the line of sight, which becomes significant in optical bands at $z>2$. We follow the prescriptions of \cite{Madau95} on the hypothesis of Ly$\alpha$, Ly$\beta$ , Ly$\gamma$ and Ly$\delta$ line blanketing, induced by H{\sc i} clouds Poisson-distributed along the line of sight.

Figure~\ref{fig:MNgal} shows examples of galaxies at redshift $z=1.6$ in the MareNostrum catalogue
 in the I, K and IRAC-8$\mu$m bands (see Appendix~\ref{sec:catalogs}) generated by the spectral synthesis described in this section.
The consistency of these colours  with classical models is checked in Appendix~\ref{s:modeldamien}.
All magnitudes are in the AB system throughout the paper.
Appendix~\ref{sec:sim-ok} carries a couple of checks
 on the simulation.
\subsection{Filaments and the skeleton}

Filaments correspond to the natural framework to characterise the environments of galaxies on  large-scale: 
within the cosmic network, large void regions are surrounded by a
 filamentary web linking haloes together.  Formally, the skeleton (\cite{skel2D,skel1}) gives a
 mathematical definition of the filaments as the locus where, starting
 from the filament type saddle points (\emph{i.e.}  those where only one
 eigenvalue of the Hessian is positive), one reaches a local maximum
 of the field by following the gradient. This involves solving
 the equation:
\begin{equation}
\frac{d \mathbf{x}}{dt}\equiv \mathbf{ v}=\nabla\rho\,,\label{eq:skel}
\end{equation}
for $\mathbf x$, where $\rho(\mathbf{x})$ is the dark matter density field,
$\nabla\rho$ its gradient, and $\mathbf x$ the position. 
Finding the large-scale structure network of filaments involves solving  Equation (\ref{eq:skel}),
 a procedure which has  recently 
been applied to both data \citep{skel2} and simulations \citep{caucci}.
A similar approach is to classify the different structures (halos, filaments, sheets, voids) according to the eigenvalues of the Hessian of the density field \citep{aragon} or the potential \citep{Pogo98, hahn,forero}.

\begin{figure}
\begin{center}
\includegraphics[width=0.95\columnwidth]{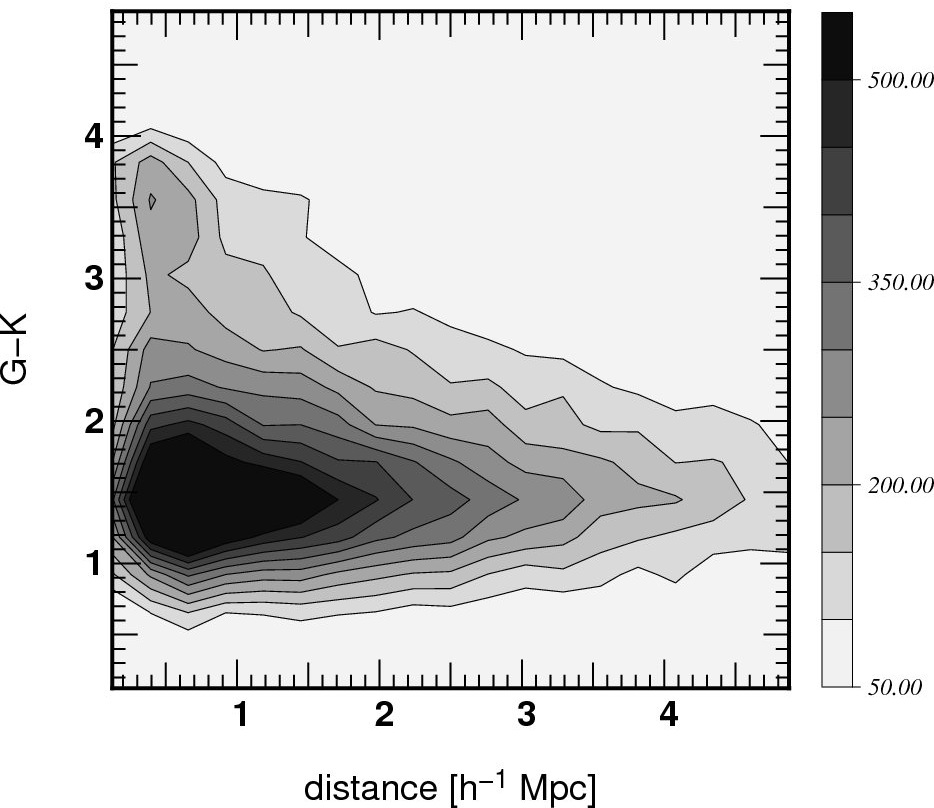}
\caption{Isocontours of the number counts of galaxies in observed colour versus distance to filament space at redshift $z=1.6$. A population of redder galaxies is present  in the close vicinity of  the filaments.}
\label{fig:color_2D}
\end{center}
\end{figure}

Recently,  \cite{sousb08} introduced  a probabilist formulation of the  skeleton,  which amounts to 
finding the solution to Equation (\ref{eq:skel}) at the intersection of the void-patches of the density field. The filaments can then be seen as the frontiers between the voids.
This algorithm has the nice feature of constructing a fully connected network of critical lines, a crucial feature for this project.
The implementation of this algorithm on the density field of MareNostrum, with a smoothing\footnote{see appendix \ref{s:smoothing} for a discussion on the effect of smoothing on the results presented here} over 2 $h^{-1}$Mpc yields a hierarchical set of 
segments, where each skeleton segment tracks its connection to its  neighbouring critical points, together with information 
relative to the underlying field (density, temperature, etc). The result is illustrated on Figure~\ref{fig:skel} where one can see that the skeleton smoothed on these scales traces well the large-scale overdense filaments, visible by eye.

\section{The influence of filaments}
\label{sec:fil}
\begin{figure}
\begin{center}
\includegraphics[width=0.85\columnwidth]{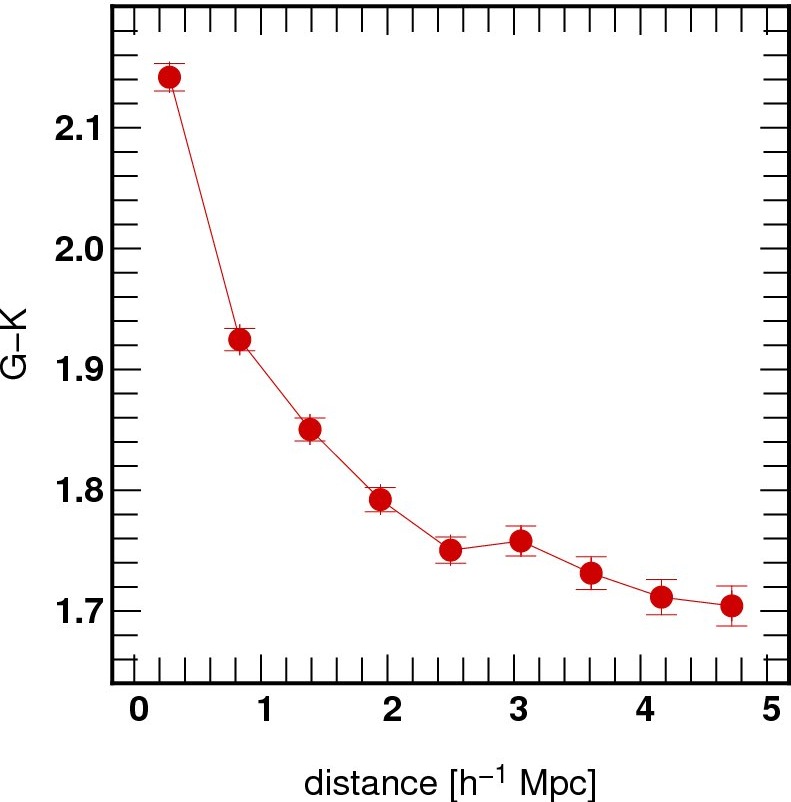}
\caption{Average observed colour as a function of distance to the closest filament at $z=1.6$. The error bars, for all the figures, represent the one sigma error on the average. Galaxies near filaments tend to be redder on average than galaxies located a few megaparsecs away from filaments.}
\label{fig:color}
\end{center}
\end{figure}

The combination of the MareNostrum simulation with spectral synthesis and the skeleton algorithm allows us to investigate the \emph{geometric} dependence of the spectroscopic properties of galaxies on the filamentary environment.

\subsection{Gradient of physical properties}
To investigate the influence of filaments on the properties of the galaxies, we choose to study the colours in the observer frame as a function of the distance to filaments. The locus of the filaments (shown in Figure~\ref{fig:skel}) is computed with the above-described skeleton algorithm, using a $256^3$ grid and smoothing on $\sigma=12$ pixels (\emph{i.e.} 2 $h^{-1}$ Mpc).
Figure \ref{fig:color_2D} represents the distribution of the observed colour G-K (which brackets the 4000~\AA{} break in the SED at this redshift) as a function of distance to large-scale filaments. First, independently to the distance to filaments, the distribution in colour shows a bimodality, that will be investigated in further details in Section \ref{sec:bimodality} : a distinct population of very red galaxies is present. Then it also shows that galaxies tend to be redder near filaments.  The trend is clearly seen when the distribution is averaged (Figure \ref{fig:color}): the G-K colour drops from 2.1 near filaments to 1.7 at a distance of 5 $h^{-1}$ Mpc. Galaxies exhibit a clear gradient of colour versus the distance to filaments.

\subsection{On the influence of nodes and filaments}
The interpretation of this gradient  is not straightforward since the distance to the skeleton does not  only reflect the influence of filaments. Indeed, clusters, located at the nodes of the skeleton, have been known  (\emph{e.g.} \cite{goto} and references therein)
 to have a strong influence on the properties of the galaxies. Galaxies near filaments are also geometrically systematically closer to nodes, and this bias could explain the observed gradient.

\begin{figure}
\begin{center}
\includegraphics[width=0.85\columnwidth]{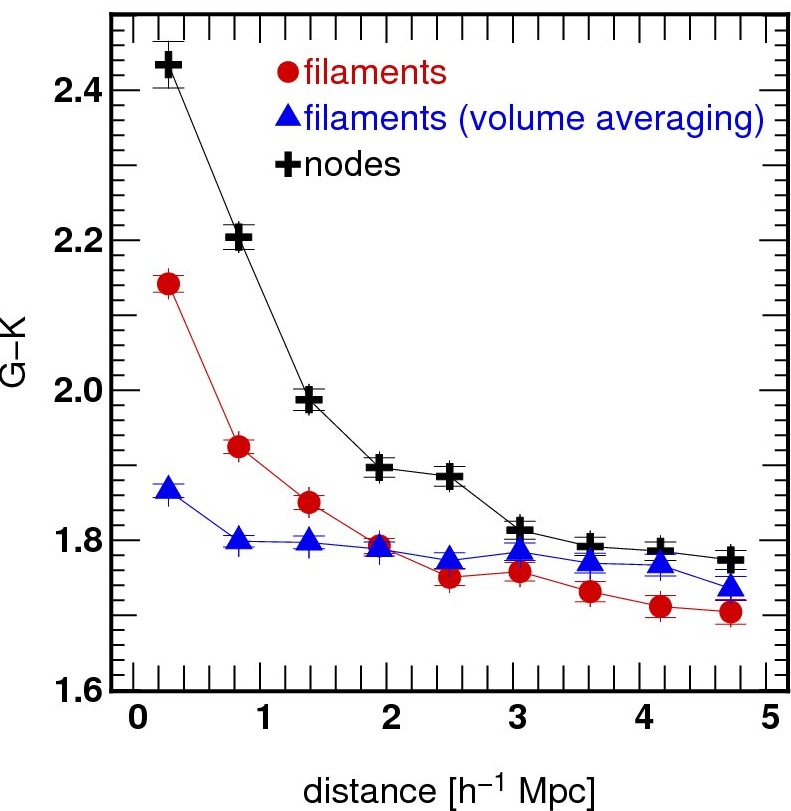}
\caption{\emph{Circles:} average colour versus distance to the closest filament at $z=1.6$. \emph{Triangles:} volume averaging. \emph{Crosses:} distance to nodes. The gradient of colour as a function of distance to filaments can be mostly explained by the sole influence of the nodes and is strongly damped when focusing on the filaments.}
\label{fig:color_all}
\end{center}
\end{figure}

 The same procedure can be applied to the distance to the nodes  alone (Figure \ref{fig:color_all}). The influence of nodes  on the colours  turns out to be even greater than the effect of the distance to filaments; it may thus explain a major part of the dependence of the colour with the distance to filaments.

One can however decrease relatively  the contribution of the nodes by using a volume average rather than a number average:
\begin{equation}
\langle x \rangle_{\rm vol} = \sum_i \frac{1}{\rho_i} x_i \left/ \sum_i \frac{1}{\rho_i} \right., 
\end{equation} where $\rho_i$ is the density of galaxies at the position of the $i^{th}$ galaxy.
Nodes contains most of the galaxies and are therefore over-represented with number averaged weighting, while volume weighted means will shift the focus on the filaments, which span on much greater scales than clusters. The volume averaged colour is plotted against the distance to the filaments in Figure \ref{fig:color_all}. The colour gradient is strongly damped, showing that most of it can be explained by the bias corresponding to the distance to nodes.

\begin{figure}
\begin{center}
\includegraphics[width=0.6\columnwidth]{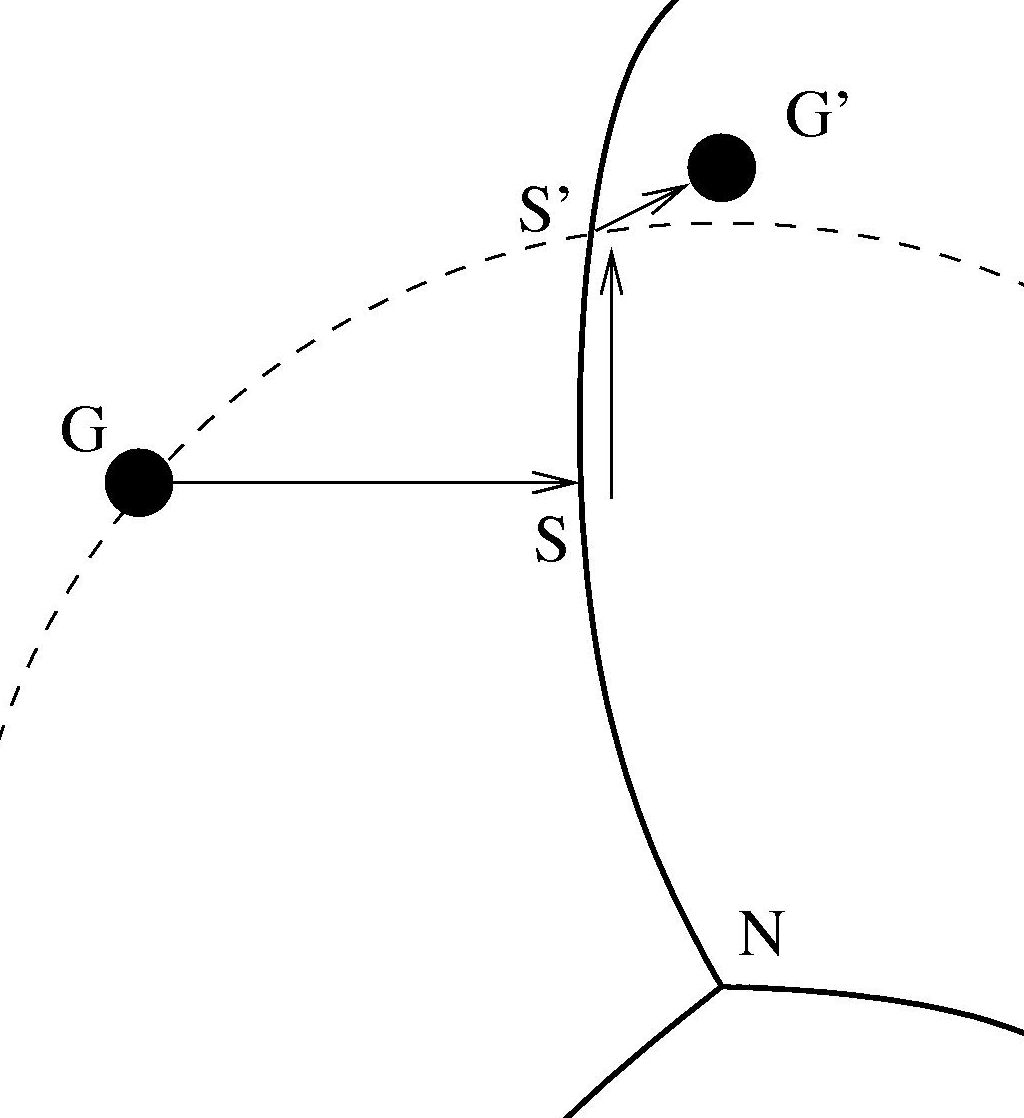}
\caption{Illustration of the algorithm used to find the filamentary counterparts. For each galaxy $G$, we find the closest skeleton segment $S$. We follow the skeleton until the point $S'$ where the distance to the node $N$ is sufficiently close to the distance between $G$ and $N$. Then the closest galaxy $G'$ is found and labelled as the filament counterpart of $G$. A galaxy and its filament counterpart have approximatively the same distance to the node, but the filament counterpart is generally much closer to the filament. Thus it opens the possibility of unbiased comparison.}
\label{fig:schema}
\end{center}
\end{figure}

\begin{figure}
\begin{center}
\includegraphics[width=0.95\columnwidth]{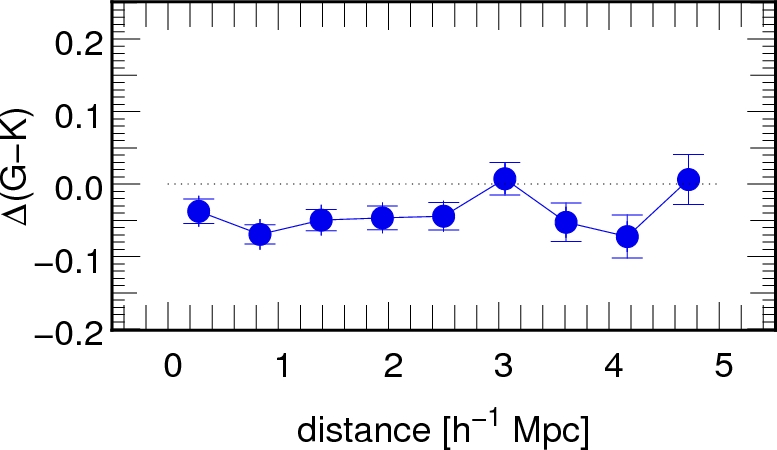}
\caption{Difference of observed colour between the galaxies and their filament counterparts versus distance to the closest filament. Even though a residual 2 sigma shift from zero remains, no overall gradient is present. Galaxies far from filaments do not tend to be different from their filament counterpart.}
\label{fig:color_pairs}
\end{center}
\end{figure}

Even if the influence of nodes is greatly reduced by volume averaging, it does not totally vanish;
 it makes it difficult to rule out a weak influence of filaments relative to a residual influence of nodes. One would want to know how the properties of a galaxy would be modified if its distance to filaments was changed while all other parameters, including the distance to nodes, are kept unchanged. A way to evaluate this is to look at galaxies having the same distance to nodes but different distances to filaments.
In order to find such pairs of galaxies, we proceed in steps (Figure \ref{fig:schema}). For each galaxy, we first look at the closest skeleton segment ($G\triangleright S$). This segment being closer to nodes, we follow the filament until we reach a segment with a distance to the node sufficiently close to the distance between the node and the initial galaxy ($S\triangleright N \triangleright S'$). The closest galaxy to this segment is considered to be the filament counterpart of the initial galaxy ($S' \triangleright G'$): it has roughly the same distance to the node, but is generally much closer to the filament. The comparison of a galaxy and its filament counterpart is therefore of much interest to study the influence of the sole filaments.
Note that this construction is not always possible. Indeed if two nodes are quite close, and if the intial galaxy is far from the filament, the filament linking them is too short to find a segment equating the distance to the node. Therefore some galaxies do not have filament counterparts and are rejected. If we accept only the galaxies which counterpart has a distance nodes greater than 90\% of the distance of the initial galaxy, around two thirds of the galaxies have a filament counterpart. Note however that the galaxies rejected are, as explained, far from small filaments, while the galaxies close to important filaments, which are of interest in our investigation, should not be affected.
Figure \ref{fig:color_pairs} shows the difference of colour between a galaxy and its counterpart. When this procedure is implemented  no significant statistical difference is found, showing that the influence of the filaments is too weak to be detected by this method and that the small gradient exhibited by the volume averaging procedure can be for the most part explained as a residual influence of the nodes.

\subsection{Other physical tracers}
\begin{figure*}
\begin{center}
\includegraphics[width=0.66\columnwidth]{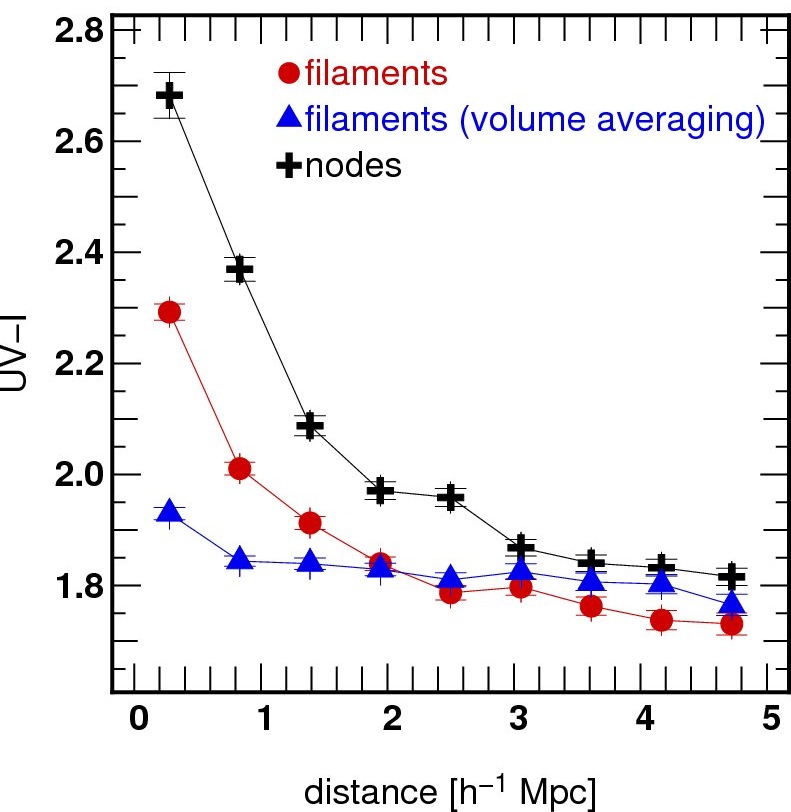} \hfill
\includegraphics[width=0.66\columnwidth]{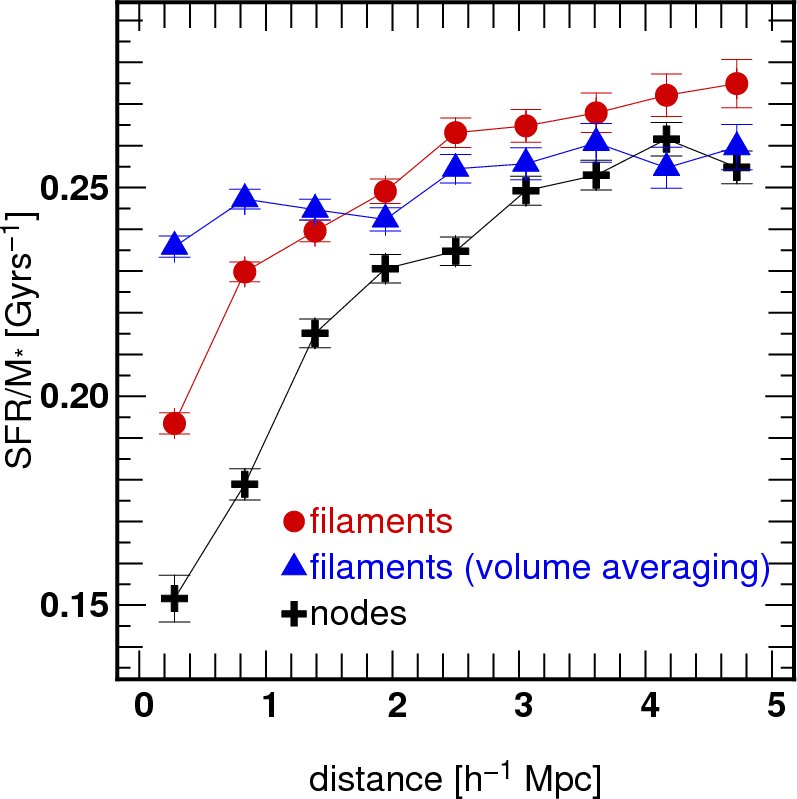} \hfill
\includegraphics[width=0.66\columnwidth]{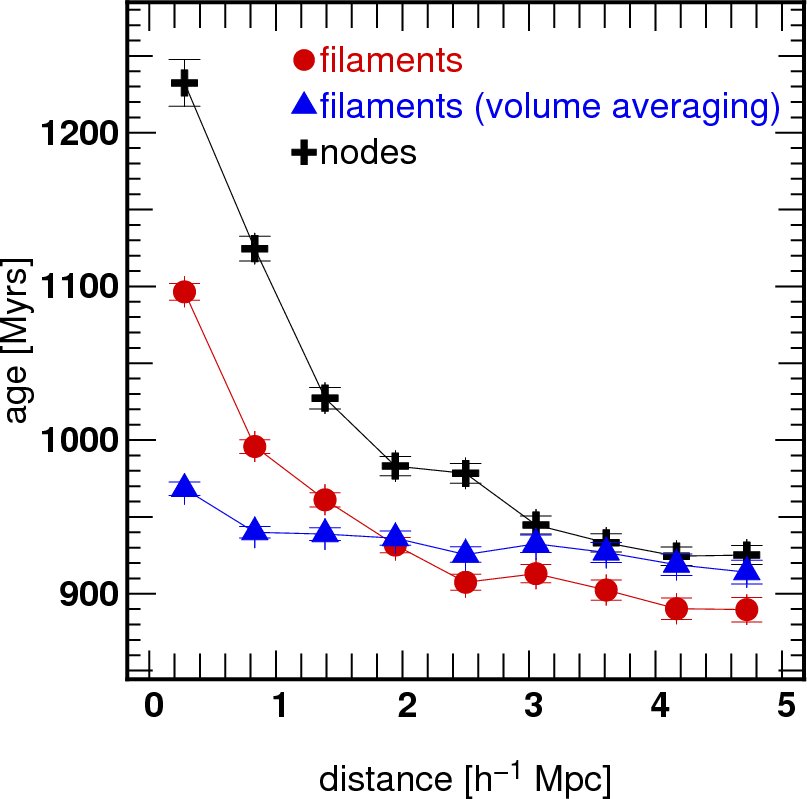}
\caption{Same as Figure \ref{fig:color_all} for UV-I rest-frame colour (\emph{left}), specific star formation rate (\emph{middle}) and mean light-weighted stellar age (\emph{right}). As expected, the same trends are found for these physical tracers as for the intrinsic colours.}
\label{fig:others}
\end{center}
\end{figure*}

The physical properties studied in the previous section are the observed colours, since  they are easy tracers to observe, and are known to reflect well the other properties of the galaxies. For the MareNostrum simulation, the same procedure can however be applied to other physical intrinsic features of galaxies. 
Figure \ref{fig:others} shows the trend for the rest-frame colour UV-I, the specific star-formation rate SFR$/M_*$ and the galaxy mean stellar age. As for the observed colour, no statistically significant gradient remains once the influence of nodes is properly removed.

\begin{figure*}
\begin{center}
\includegraphics[width=0.95\columnwidth]{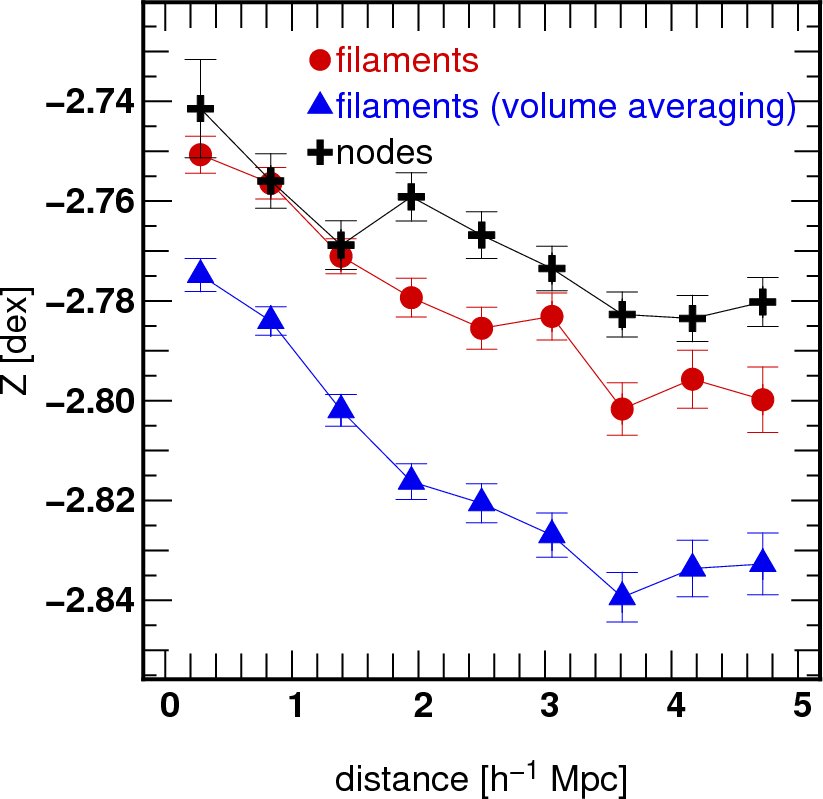}\hfill
\includegraphics[width=0.925\columnwidth]{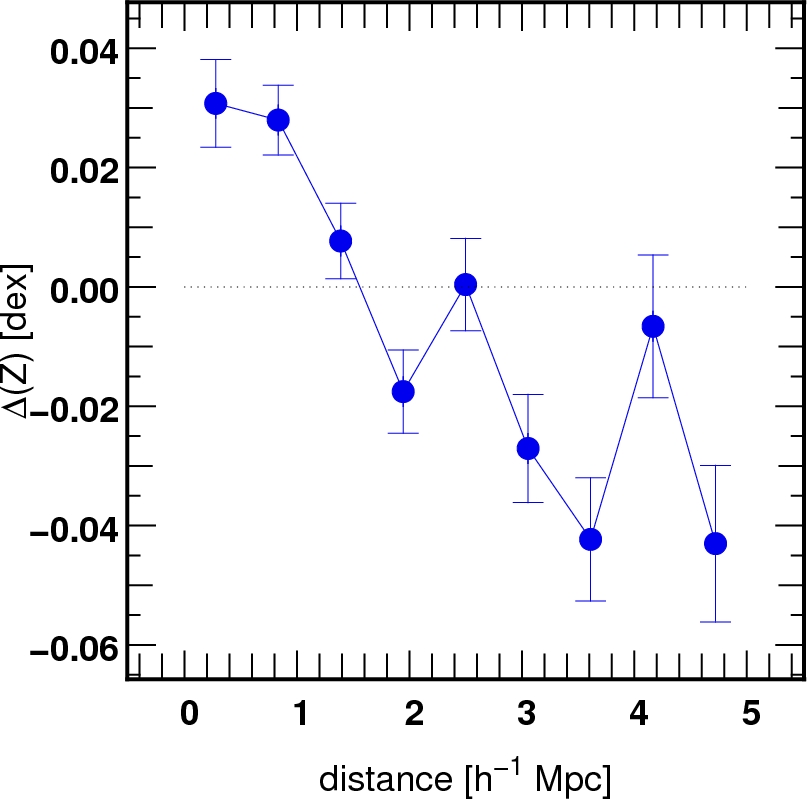}
\caption{Same as Figure \ref{fig:color_all} \emph{(left)} and Figure \ref{fig:color_pairs} \emph{(right)} for  the metallicity. Metallicity shows a clear gradient as a fuction of distance to filaments. This gradient reflects the inhomogeneity of the WHIM on these scales for the MareNostrum subgrid physics.}
\label{fig:metal}
\end{center}
\end{figure*}
Figure \ref{fig:metal} shows the trend for the metallicity. A gradient is still present after volume averaging and can be seen in pair comparison. The filaments seem to have an influence on the metallicity of the galaxies: the closer to the filaments the galaxies are, the more metallic they are. This gradient is however fairly small (0.1 dex is  the order of magnitude of the global change in the metallicity of galaxies within 1 Gyr at this epoch, see \cite{Savaglioetal05}), which is consistent with the fact that the other properties, which are indirectly related to metallicity, do not exhibit any significant gradient.
This gradient may nevertheless reflect the enrichment of the Warm-Hot Intergalactic Medium (WHIM,\cite{WHIM2}). 
Indeed, it is a consequence of the increased galaxy density, together with 
the corresponding  (in-)efficiency of winds as induced by the subgrid physics.
If this result were confirmed not to depend critically on  the  supernovae rate recipe,
it offers the prospect of directly exploring this component of the IGM via the metallicity of galaxies. It would then be of interest to cross correlate this metallicity with that 
of the O$_{\rm IV}$ absorbers \citep{WHIM}.

\section{Bimodality within clusters}
\label{sec:bimodality}

\begin{figure}
\begin{center}
 \includegraphics[width= 0.85\columnwidth]{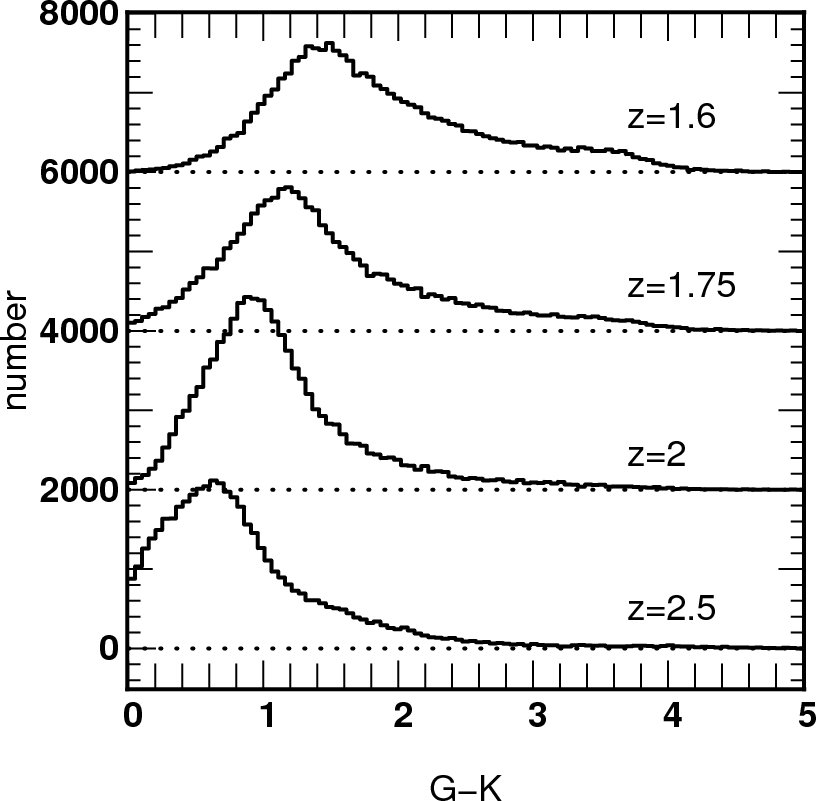}
\caption{Distribution of observed colour for several redshifts ($z=1.6, 1.75, 2.0$ and $2.5$). The curves are shifted by 2000 for clarity.
Note the weak bimodality occurring  below redshift 2.}
\label{fig:color_histo}
\end{center}
\end{figure}

In addition to its evolution with the distance to filaments, the presence of a bimodality corresponds to an interesting feature of the distribution of colours: as seen on Figure \ref{fig:color_2D}, a population of very red galaxies ($G-K>3$) is present. Moreover, the evolution of the colour distribution with redshift shows that this bimodality is appearing around $z=2$ (Figure \ref{fig:color_histo}).
Let us therefore trace back the positions of  galaxies responsible for this bimodality in order to explain its origin.

\subsection{Properties of the reddest galaxies}
\begin{figure}
\begin{center}
\includegraphics[width=0.85\columnwidth]{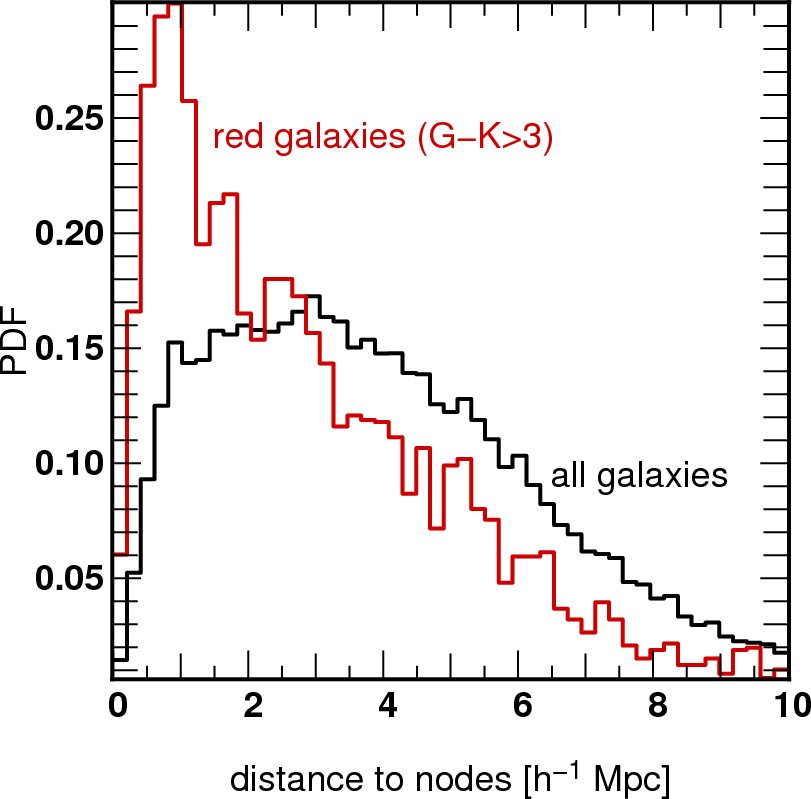}
\caption{PDF of the distance to nodes for all galaxies (\emph{black}) and reddest ($G-K>3$) galaxies (\emph{red}). The reddest galaxies are clustered near the nodes.}
\label{fig:redclustering}
\end{center}
\end{figure}

\begin{figure}
\begin{center}
\includegraphics[width=0.9\columnwidth]{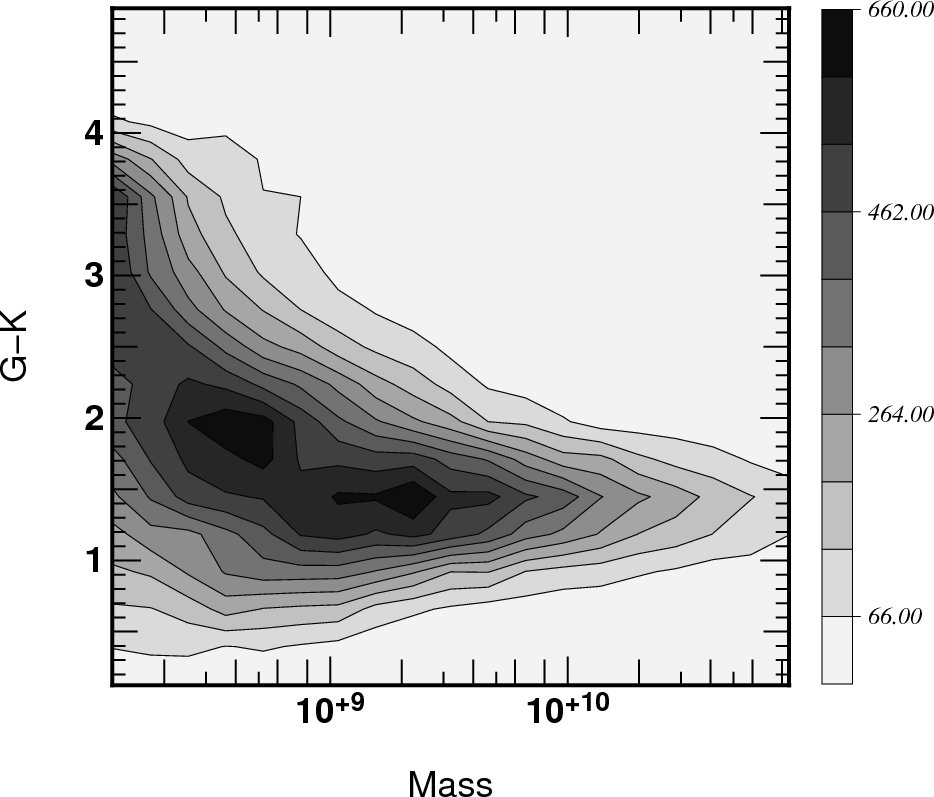}
\caption{Isocontours of  the density of galaxies  in mass versus colour space,  for $z=1.6$. Red galaxies are  typically less massive.}
\label{fig:mass}
\end{center}
\end{figure}

Figure \ref{fig:redclustering} shows the distribution of distance to nodes for the reddest galaxies ($G-K>3$) and for the overall sample. Red galaxies are shown to be more clustered near nodes.

These red galaxies also exhibit a very low mass ($M \sim 10^8 M_\odot$).  Indeed, as figure \ref{fig:mass} shows, the galaxies are either blue (with high or low masses) or red with low masses. Such small masses associated to very red colours (which in our case mean an absence of star formation since we do not include internal dust in the modelling of the SEDs) suggest that these may be dwarf spheroidals, which origin is still poorly known. \cite{Grebeletal03}, in a study of dwarf galaxies in the Local Group,  suggested that the absence of star-formation in such objects is due to externally induced gas loss: ram-pressure stripping would be responsible for lack of ISM gas in these small galaxies. \cite{Mayer} showed that stripping in the clusters can indeed increase the mass-to-light ratio of dwarf galaxies. It is possibly the case here too, as demonstrated in Sect.~\ref{section:dyn} and Figure~\ref{fig:traces}. In contrast to what is observed at low redshift, there is no massive red galaxy in the simulation at $z=1.6$. \footnote{Low mass galaxies should be considered with caution in such a simulation. For instance  low mass  galaxies are expected to  be artificially  blue as they only have a recent accretion history, whereas more massive galaxies are spectrally better resolved. However this effect would not explain the fact that the galaxies studied are redder.}

\subsection{Dynamical scenario}
\label{section:dyn}
\begin{figure*}
 \begin{center}
 \includegraphics[width=0.475\textwidth]{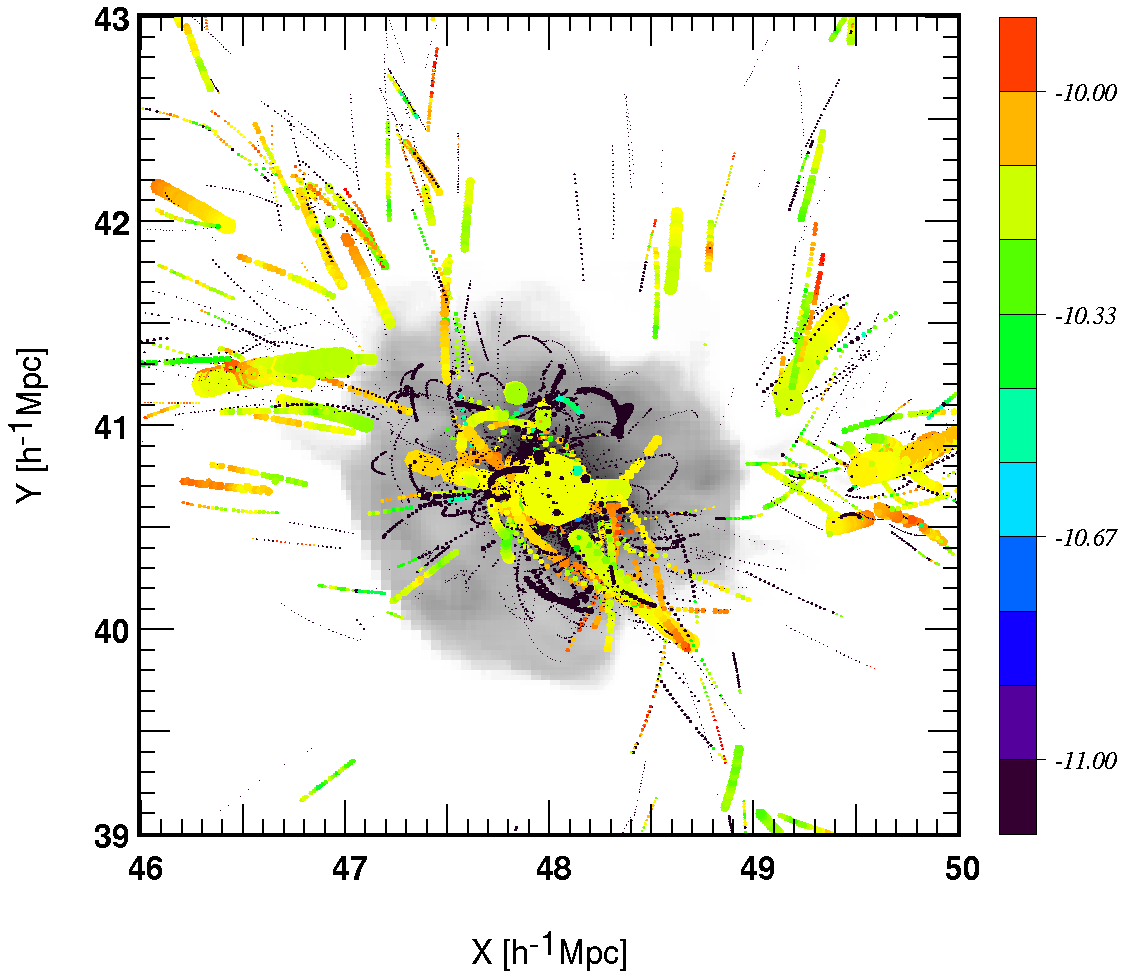}
  \includegraphics[width=0.475\textwidth]{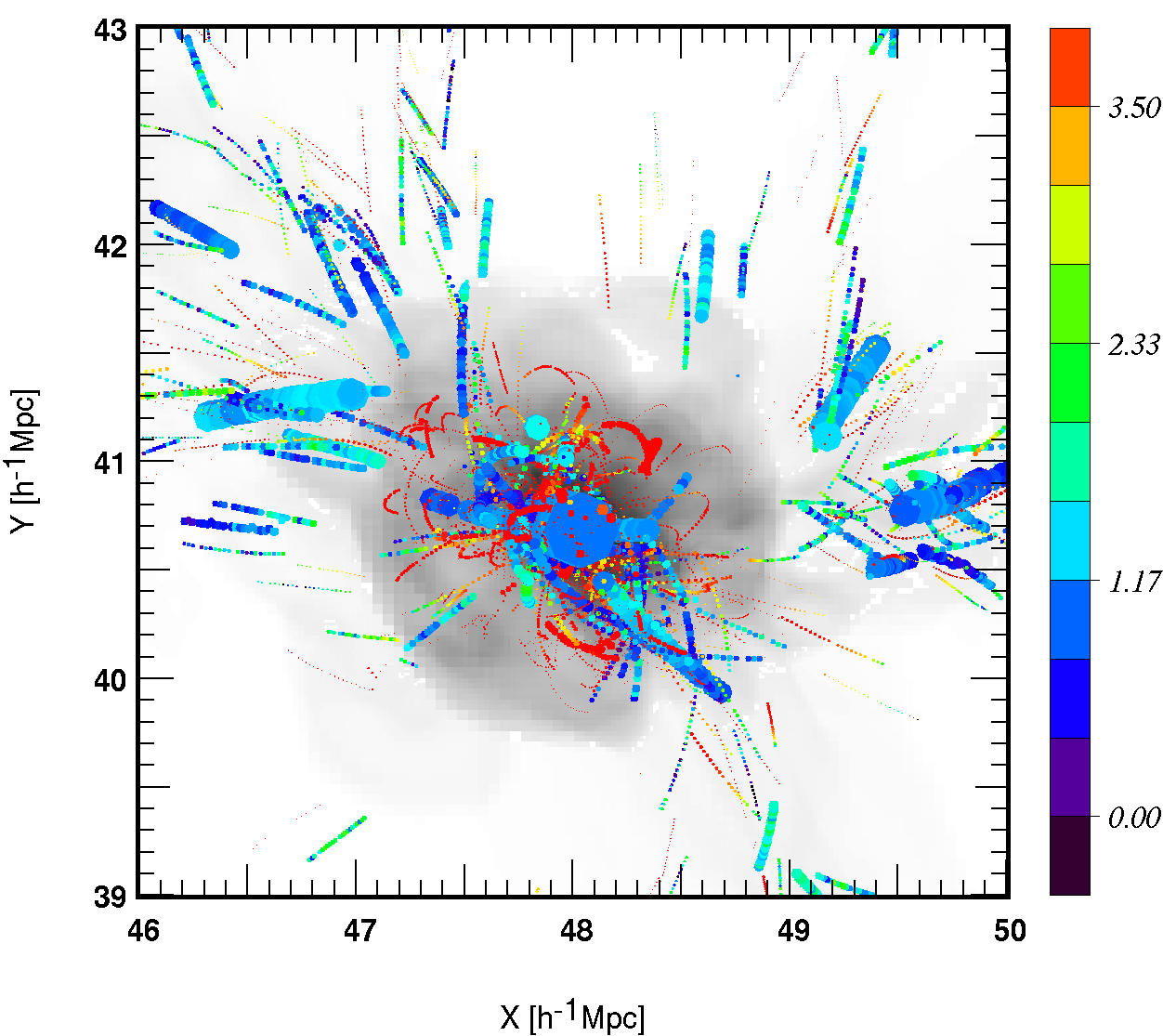}
  \includegraphics[width=0.475\textwidth]{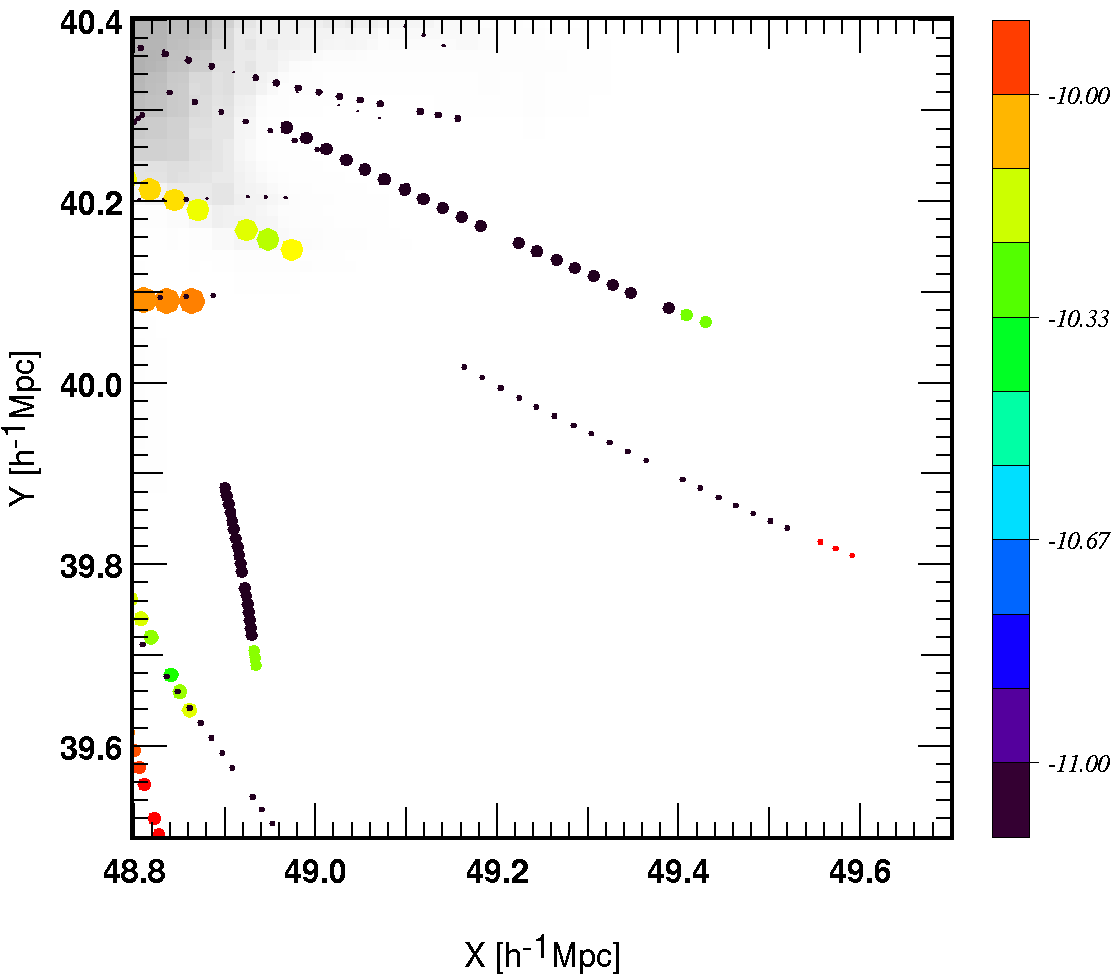}
  \includegraphics[width=0.475\textwidth]{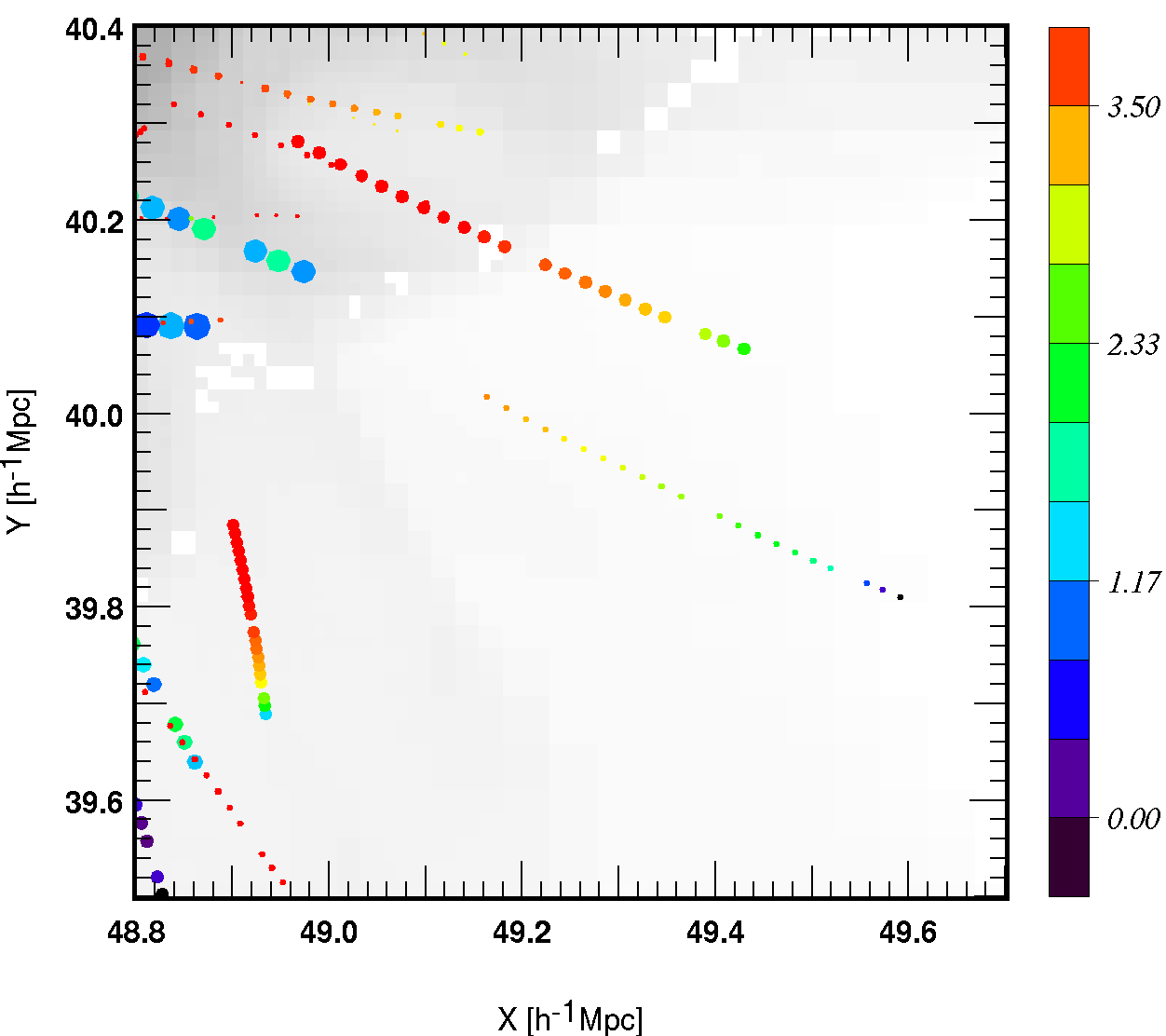}
\caption{View of a small region of the simulation centred on a large cluster. The traces represent the position of the galaxies for several timesteps between $z=1.9$ and $z=1.6$.  Each galaxy is represented by a point whose size depends on its mass and whose colour represents either the SFR$/M_*$ (\emph{left}) or the UV-I rest-frame colour (\emph{right}). The top figures represent a 4 $h^{-1}$Mpc cube and the bottom ones are a zoom on the bottom right part and span over 1 $h^{-1}$Mpc. The grey background encodes the temperature of the gas, showing the extent of the hot gas bubble. Note the reddening of the small galaxies entering the cluster ({\sl bottom right}), in parallel to the reduced star formation  ({\sl bottom left}).
}
\label{fig:traces}
 \end{center}
\end{figure*}

The dependence of the spectro-photometric properties with the distance to filaments studied in the previous sections reflects the process of galaxy formation and its connection with 
the global flow within the large-scale structure. Indeed, it has been demonstrated \citep{aubert,skel1} that filaments are fed by surrounding voids and mark the lanes of 
galactic infall towards the clusters.
Young galaxies form across the whole filamentary network, but are rapidly collected along the more busy subnet of 
denser filaments.

A close-up look at a specific halo (see Figure \ref{fig:traces}) confirms the presence of small, red galaxies (which do not form stars) embedded in the halo. These galaxies are already accreted at $z=2$. At this range of redshift, haloes contain big ($M > 10^{10} M_\odot$), blue, central galaxies and small ($M < 10^9 M_\odot$), red galaxies which are being stripped of their gas and swallowed by the central galaxy. The fate of these small galaxies can be investigated by looking to another population: the galaxies that are currently accreting into the halos at this epoch, although they have a larger mass ($10^9 M_\odot< M <  10^{10} M_\odot$).
These intermediate galaxies shows a peculiar behaviour that could explain the observed increase of the average colour near nodes: when approaching the halo, they stop forming stars and therefore begin to passively redden. 
 This quench could be triggered by the lack of cold stream reaching the inner core of dark halos \citep{ocvirk2008} or by stripping of their gas \citep{stripping}.  \cite{ocvirk2008} show that more massive haloes prevent cold stream from feeding the inner halo;
here it is found that these halos display redder satellites.
The accreted galaxies tend therefore to be red and have a low SFR, leading to the previously observed bimodality. 
They would then merge with the central galaxies, and contribute to the increase in mass (dry merging) of the massive old galaxies observed at low redshift.


\begin{figure}
 \begin{center}
 \includegraphics[width=0.85\columnwidth]{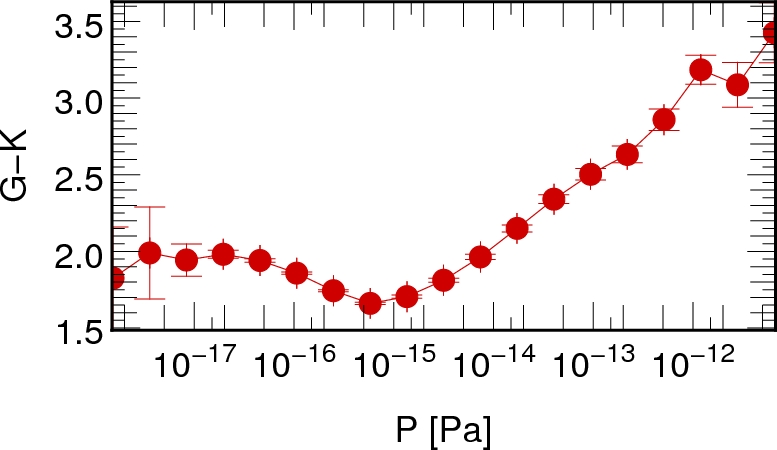}
\caption{Mean observed colour, G-K, as a function of pressure of the galaxy. Galaxies are redder in high pressure regions.}
\label{fig:color_pressure}
\end{center}
\end{figure}

\begin{figure}
 \begin{center}
  \includegraphics[width=0.85\columnwidth]{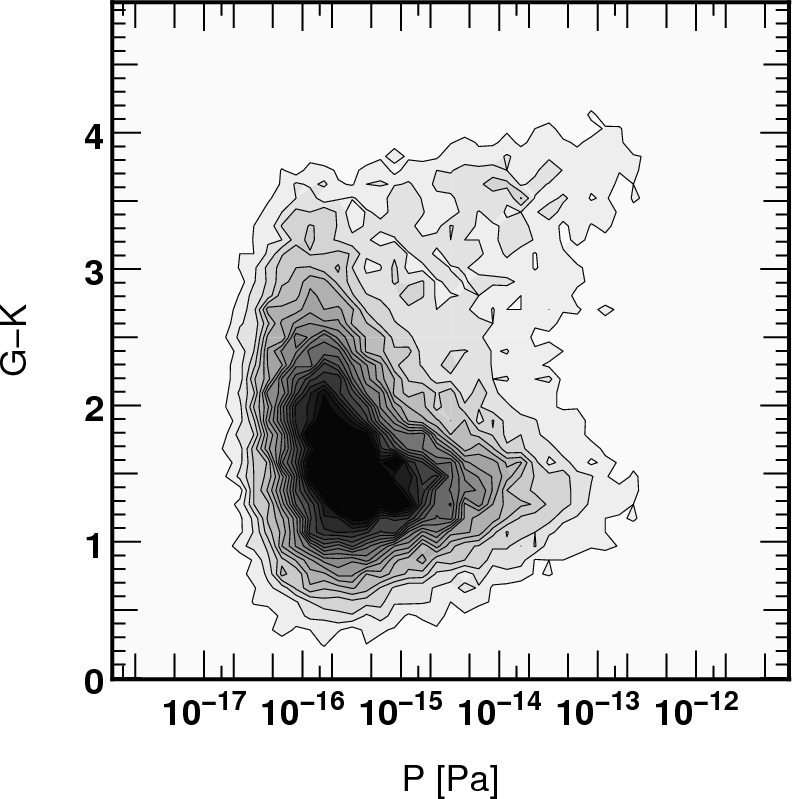}
\caption{Distribution of  galaxies in observed colour (G-K) pressure space. The increase of the G-K colour with pressure is due to a distinct population.}
\label{fig:color_pressure_2D}
 \end{center}
\end{figure}

In this scenario, the spectroscopic properties of a galaxy should be at least in part determined by the physical conditions of the intra-cluster medium.
Figures \ref{fig:color_pressure} and \ref{fig:color_pressure_2D} are consistent with this scenario. Indeed the average colour increases with the pressure, as expected in the case where the reddening is induced by  star formation quenching via ram-pressure stripping. The detailed 2D distribution shows the presence of a population of very red galaxies for high pressures. This galaxies could correspond to the galaxies which have been stripped of their gas when their entered the cluster, and will then become red, small galaxies, before they eventually merge with the core galaxies.

\begin{figure}
 \begin{center}
  \includegraphics[width=0.85\columnwidth]{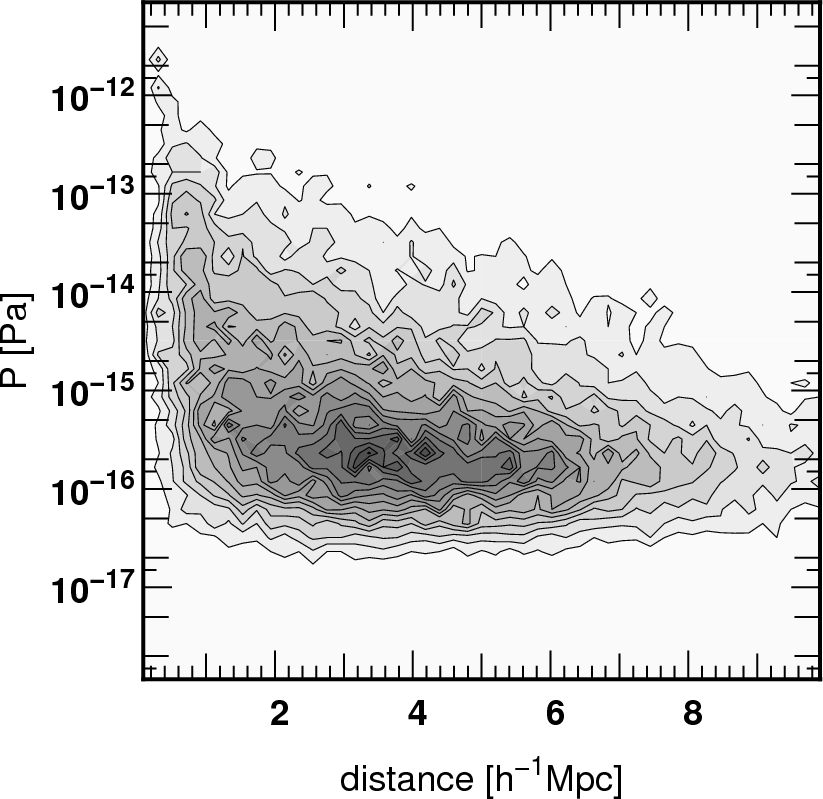}
\caption{Pressure of the galaxies versus distance to nodes. High pressures (likely to strip low mass galaxies of their gas) are only found near massive clusters.
}
\label{fig:pressure_distance}
 \end{center}
\end{figure}

The typical pressure of transition is around $10^{-14} $ Pa and corresponds to conditions found only near nodes (see Figure~\ref{fig:pressure_distance}). Theoretical considerations and simulations \citep{fujita, roediger} show that ram-pressure stripping is expected to be efficient for pressure over $10^{-13}$~Pa. 
This apparent discrepancy may be explained by noticing that we are considering different objects. Their value corresponds the pressure required to strip a typical spiral galaxy at low redshift and could thus be lower for the smaller high redshift galaxies we are considering.

\subsection{Tentative link with observations}\label{sec:discussion}

It is clearly beyond the scope of this paper to carry a direct comparison with the currently available data.
Indeed the main difficulty in comparing  the results  found in this simulation with existing observations lies in the range of redshifts and the geometry of the surveys. 
Building a fully connected skeleton is non trivial for pencil shape volumes, as edge effects become important (as discussed in \cite{skel1}). Moreover, 
 in this paper, the skeleton was computed from the dark matter distribution and one would need to calibrate the bias involved in using light instead of mass. 

Finally, the simulation was stopped at $z \approx 1.5$, and the bimodality seems to appear around $z\approx 2$ (see Figure \ref{fig:color_histo}). Observations at such high redshift are uncommon and difficult \citep{vanDokkum, Daddi2005}.  
However, as mentioned in Section~\ref{sec:intro}, similar trends exist for the same range of redshift: for example, red galaxies are observed to be more clustered \citep{Daddi2003}.

\begin{figure}
\begin{center}
\includegraphics[width= 0.95\columnwidth]{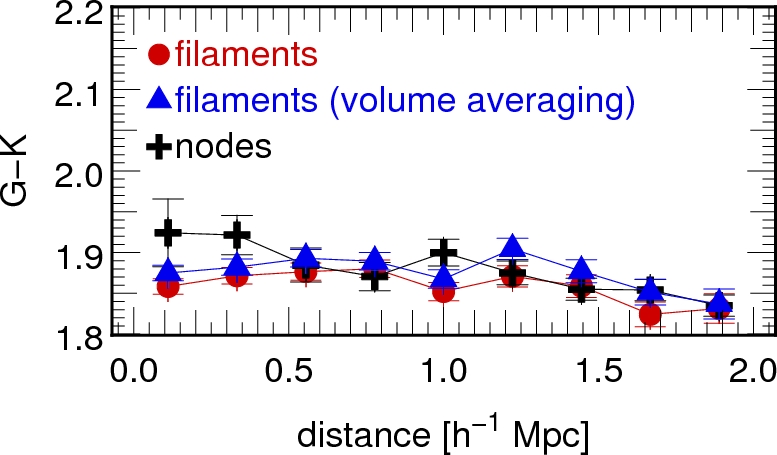}
\caption{Same as Figure \ref{fig:color_all} for the 2D skeleton. A weak trend is still present but much less significant than in 3D.}
\label{fig:2D}
\end{center}
\end{figure}

These gradients can also be linked to slightly lower redshift observations. For redshift $z\approx1$, there exists several surveys that have studied the role of the environment, \emph{e.g.} GOODS \citep{Elbaz}, VVDS \citep{Scodeggio} or DEEP2 \citep{Cooper2006}. The existence of a bimodality, with a blue and a red sequence, is known to appear at this redshift \citep{Nuijten2005} and can be observed on several properties of galaxies at low redshift \citep{Mouhcine}.\\
To better understand the link with observations, one has to understand the robustness of the results with respect to observational uncertainties and other biases.
\label{s:projobs}
The nature of this virtual data set allows us to compute directly the properties (skeleton, distance to filaments, \dots) in 3D, without any problems  of distance determination. For observational applications, a natural question arises: are these results robust with respect to distance uncertainty, that can come e.g. from the use of photometric redshifts \citep{Cooper2005}?
In order to asses  this robustness, the simulation cube is projected along one of its axis. The 50 $h^{-1}$ Mpc can be thought of  as representing the uncertainty on distance (roughly $\delta z = 0.005$ at $z=1.6$). A 2D skeleton is computed on the resulting density field, with the same N-dimensional algorithm. The observed colour is then compared to the 2D distance to the closest filament (see Figure \ref{fig:2D}). The main features found in the 3D investigation are still weakly present after projection: a distinct population can be seen near nodes and leads to a enhanced averaged colour, but filaments do not seem to have an effect on the properties of the galaxies. This effect is however far more subtle than in 3D, given the projection effect.

\begin{figure}
\begin{center}
\includegraphics[width= 0.95\columnwidth]{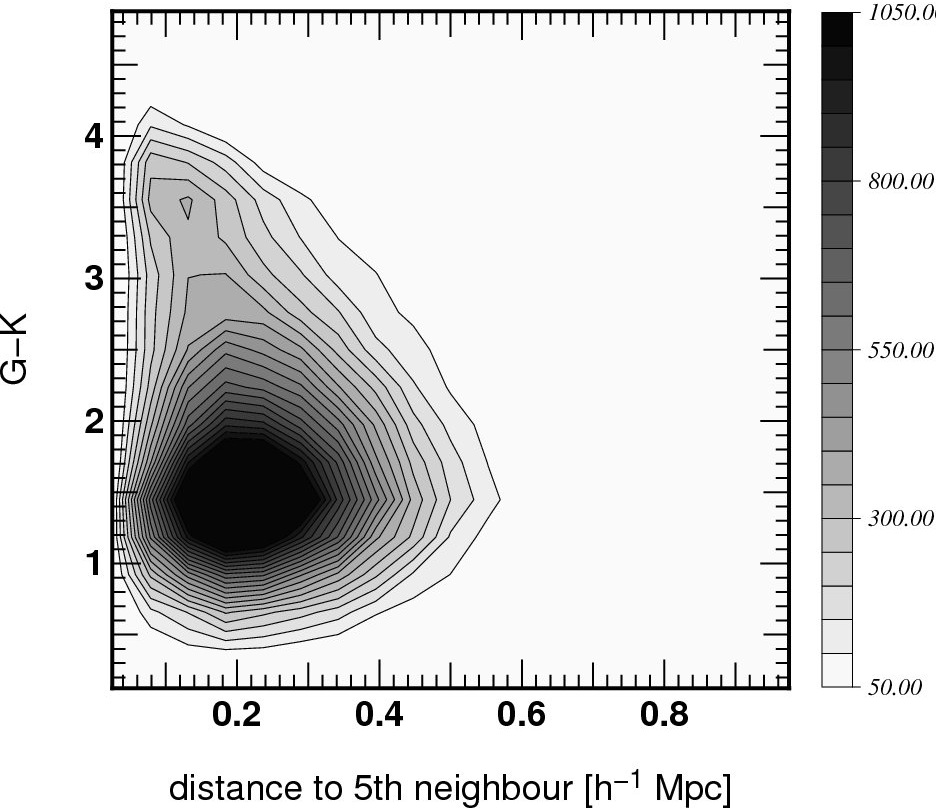}
\caption{Isocontours of the number count  of galaxies in observed colour versus distance to the 5$^{\rm th}$ neighbour for the projected data
 at $z=1.6$. Note the larger number of contours used to catch the low contrast bimodality, compared to Figure \ref{fig:color_2D}.
 Note also the difference of scale on the $x$ axis: the $5^{\rm th}$ neighbour does not probe large-scale structures when no cut on galactic mass is applied. }
\label{fig:d5}
\end{center}
\end{figure}

Finally, as an alternative to the skeleton as a tracer of filaments 
let us briefly implement a few more commonly used  tracers (as mentioned in the introduction) on the MareNostrum simulation: the 2D distance to the 5$^{\rm th}$ neighbour \citep{Cooper2005} and the galaxy number density. Figure \ref{fig:d5} shows that a bimodality is still present with the 5$^{\rm th}$ neighbour probe,
 but it is much less contrasted  and much more localised. 
The 2D distance to the 5$^{\rm th}$ neighbour is therefore a less sensitive probe of the anisotropic cosmic environment.
 In contrast, 
provided spectroscopic redshifts are available (say with the LSST, \cite{LSST}), the 3D skeleton should allow us both to probe the large 
scale structures and mark the neighbourhood of clusters in detail.

\begin{figure}
 \begin{center}
  \includegraphics[width=0.95\columnwidth]{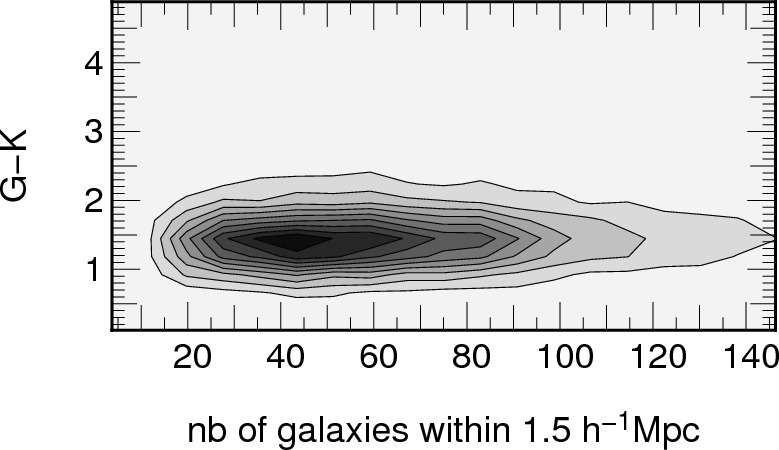}

  \includegraphics[width=0.95\columnwidth]{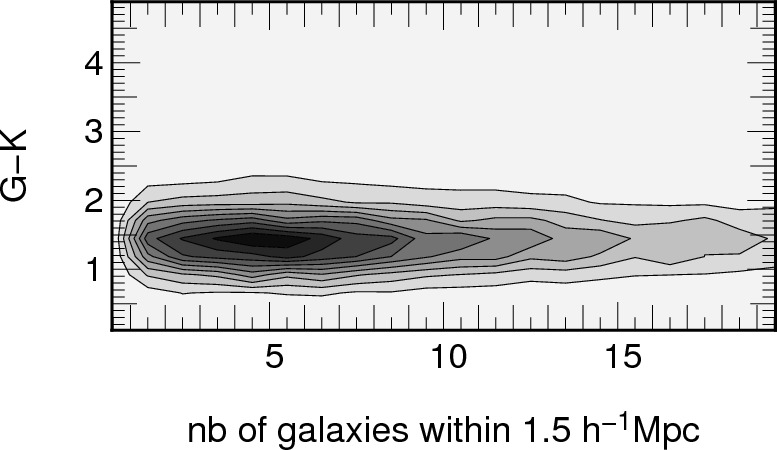}
\caption{Isocontours of the number count of galaxies in observed colour versus number of galaxies within a sphere of 1.5 $h^{-1}$Mpc for the projected data (\emph{top}) and the 3D data (\emph{bottom}). No significant trend is seen with these classical tracers.}
\label{fig:nbvoisins}
 \end{center}

\end{figure}

Finally, figure \ref{fig:nbvoisins} shows that the bimodality is not present anymore when the galaxy number density is used, in 2 or 3D.

\section{Conclusion \& Discussions}

\label{sec:conclusion}

The Cosmic Web is a key feature of the organisation of galaxies on large scales. This paper investigated the influence of this filamentary environment on the spectroscopic properties of galaxies, using the MareNostrum simulation which was postprocessed using stellar population synthesis. 
The cosmic web was traced  with the skeleton algorithm, and  has proven here 
to be a very effective mean of probing the anisotropy of the large-scale structures.

We found gradients of spectroscopic properties of galaxies with the distance to filaments, but demonstrated that they can be explained by the fact that the distance to filaments is biased by the distance to the nodes of the network (group or clusters of galaxies). 
Two procedures were introduced to remove the influence of these nodes: (i) volume-averaging  this distance decreases the influence of the compact, dense regions and focus on wider structures, such as filaments; (ii) pair comparison which seeks the filamentary counterpart of each galaxy. Both methods show that the influence of the filaments alone is negligible compared to influence of clusters.
A bimodality in colour was also found to occur below redshift $z\approx 2$ and its origin was investigated. It  is due to a population of red, small galaxies ($\sim 10^8 M_\odot$) accreting on the nodes of the Cosmic Web, 
while more massive objects, $M> 10^8 M_\odot$, are mostly unaffected. These galaxies have their star formation quenched while they approach the clusters.
It remains to be confirmed that this stripping process is not amplified by a lack-of-resolution effect, since (i) it involves  amongst the smallest (virtual) galaxies in the simulation, and (ii) it seem to create a tension with observations 
at redshift zero \citep{kimm}.

These findings suggest that the large-scale filaments are only dynamical features of the density field, reflecting the flow of galaxies accreting on clusters; the conditions in the filaments are not dramatic enough to influence strongly the properties of the galaxies it encompases, unlike the intra (proto) cluster medium, which seems able to strip down the ISM of the  incoming low mass galaxies. Appendix \ref{sec:test_cut} shows that  this statement remains valid when the study is limited to the most important filaments.

Finally, we did  find a weak metallicity gradient away from the filaments which could reflect the large-scale inhomogeneity in the distribution of metals (the so called WHIM).

One could have imagined that even if the  large-scale filamentary network were  purely a tracer of the large-scale dynamics, 
galaxies within the large-scale filaments should be redder and older, since they would have joined the cosmic super highway earlier on average when compared to field galaxies. This effect is not seen at those redshifts, as (i) older galaxies continue to accrete new cold gas on small scales and form stars, hence remain blue, (ii) some field galaxies continuously join the large filamentary network, and (iii) on large-scales, a significant fraction of the matter is accreted more or less radially onto the nodes, while filaments are  collecting  left-overs from this accretion more indirectly \citep{skel1}.

Recently \citep{keres05,ocvirk2008,dekelnature},  it was emphasized in steps that anisotropic metal-rich cold stream accretion regulates the inflow of cold gas towards the inner regions of 
the most massive galaxies of the high redshift universe ($z>2$). 
It was then conjectured that this process could explain  the observed  bimodality of spectroscopic galactic properties at
lower redshift.  In this paper, we have shown that the geometric distribution of the colour of galaxies is not sensitive to the detailed large-scale filamentary network, but only to its nodes. 
 This apparent paradox may be lifted when noting that the self-regulating anisotropic filamentary accretion  occurs on much smaller scales, and was quantified for the central      galaxies of the simulation which {\sl are} sitting at the nodes of the network; in contrast, when considering the full galactic population, a typically low mass galaxy
 is not transformed by its encounter with the different physical condition of the weakly overdense intergalactic medium within the cosmic web, unless it falls into the intra cluster
 medium of a large node.\\
 In other words, massive galaxies feel the small scale filaments (cold streams) feeding them at nodes; 
low mass galaxies are not spectroscopically changed while entering the large-scale filaments. 
The mesoscopic (below a Mpc scale) filamentary feature of the cosmic network
may geometrically solve the self-regulating process of galactic accretion, but  we have demonstrated here that 
its large-scale counterpart does not seem to directly affect the colours of galaxies. 

In \cite{sousb08}, the effect of redshift distortion is partially addressed, and the corresponding  algorithm is now being 
extended to discrete surveys via a Delaunay tessellation (which are therefore not sensitive to edge effects).
This, as argued in Section~\ref{sec:discussion} is a critical step towards performing a similar 3D analysis on real data,
which as we have shown is essential to quantify these gradients, as in projection, the information is lost (see Figure~\ref{fig:2D}).
In particular, it would  be of great interest to carry out these measurements on a  DEEP-2/VVDS/z-COSMOS-like survey as 
well as 
 to bring the simulation down to lower redshift and reach a time in cosmic history when upcoming large  observational surveys (\emph{e.g.} LSST \cite{LSST}, BOSS \cite{BOSS}) overlap 
statistically with the predictions of the simulation.  It would also be worth investigating how sensitive some of our findings are w.r.t. the detailed chosen subgrid physics by probing 
alternative recipes and running higher  spatial resolution simulations.

The catalogues produced for this investigation (spectroscopical properties of the MareNostrum galaxies and its dark matter skeletons)
are available online as discussed briefly in Appendix~\ref{sec:catalogs}. 

\subsection*{Acknowledgements}
{ \sl
We thank F.~Brault for her help in  a preliminary investigation,  S.~Colombi, D.~Pogosyan, Y.~Dubois and
D.~Aubert  for fruitful
comments during the course of this work. This investigation was
carried within the framework of the Horizon project,
\texttt{www.projet-horizon.fr}.  
  The simulation was run on the MareNostrum machine at the Barcelona Supercomputing 
  Centre and we would like to warmly thank the staff for their support and 
  hospitality. We  also thank D.~Munro for freely
distributing his Yorick programming language and opengl interface
(available at {\em\tt http://yorick.sourceforge.net/}). 
CP thanks the  Leverhulme Trust for the visiting professorship F09846D.
}

\bibliographystyle{mn2e}
\bibliography{filaments}

\appendix

\section{Self-consistency checks}
In this section, we present a few tests that were performed to check the presence of bias and the influence of the parameters of the skeleton and the MareNostrum simulation.

\subsection{Decorrelating galaxies with their environment}

\begin{figure}
\begin{center}
\includegraphics[width=0.85\columnwidth]{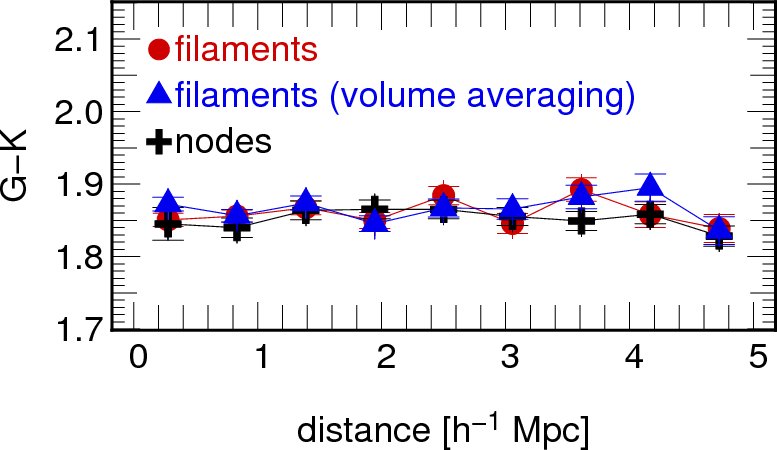}
\includegraphics[width=0.85\columnwidth]{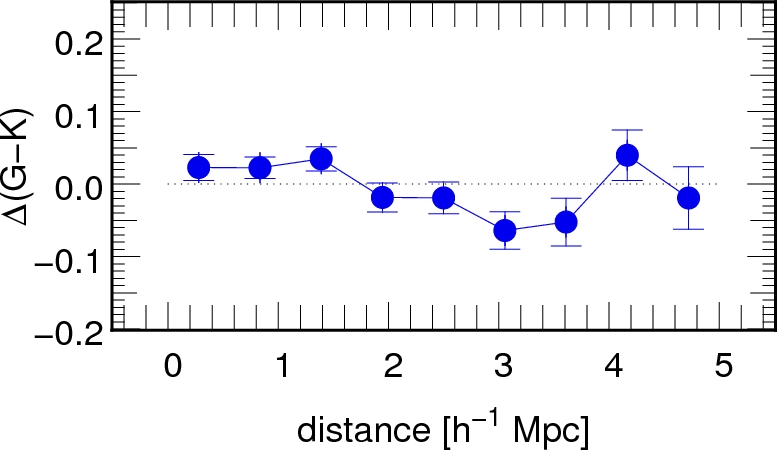}
\label{fig:test_aleat}
\caption{Same as Figure \ref{fig:color_all} (\emph{top}) and Figure \ref{fig:color_pairs} (\emph{bottom}) when galaxies properties are shuffled. As expected, the evolution is canceled.}
\end{center}
\end{figure}

In order to check if our method induces artificial correlations
 of the galaxies properties with their environment, we
 shuffle the properties of the galaxies: for each galaxy, we change its colour to the colour of another randomly chosen galaxy, 
 keeping its position unaffected. The results obtained on the shuffled sample, corresponding to  Figure~\ref{fig:color_all}, are presented in Figure~\ref{fig:test_aleat}. As expected, the evolution of colour with the distance to filaments or nodes is totally erased.

\subsection{The influence of smoothing}
\label{s:smoothing}
\begin{figure*}
 \begin{center}
  \includegraphics[width=0.3\textwidth]{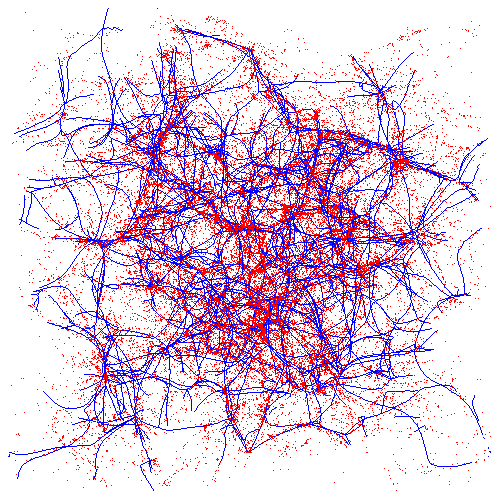} \hfill
  \includegraphics[width=0.3\textwidth]{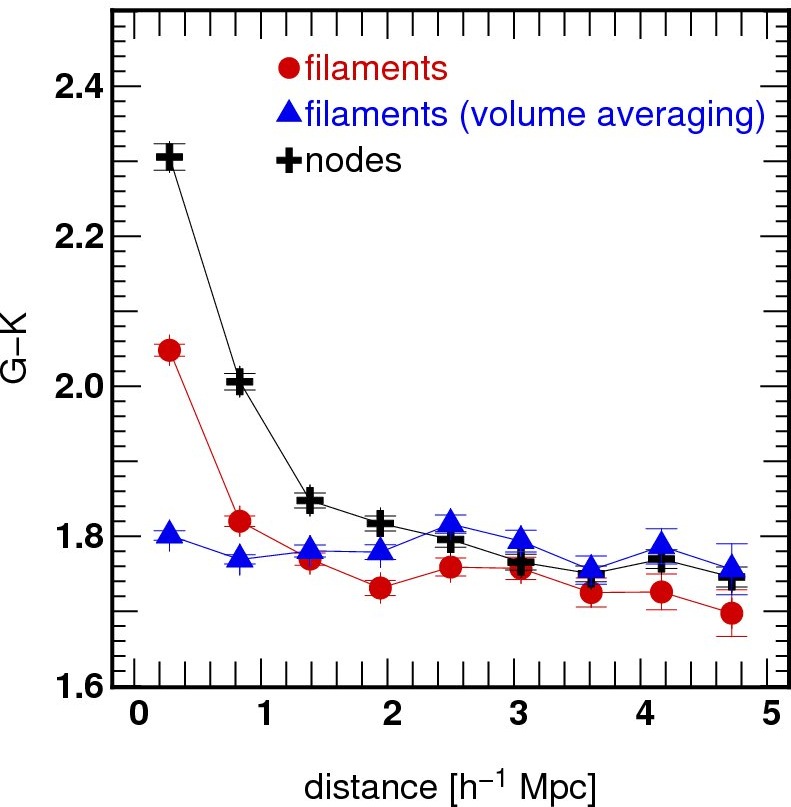} \hfill
  \includegraphics[width=0.3\textwidth]{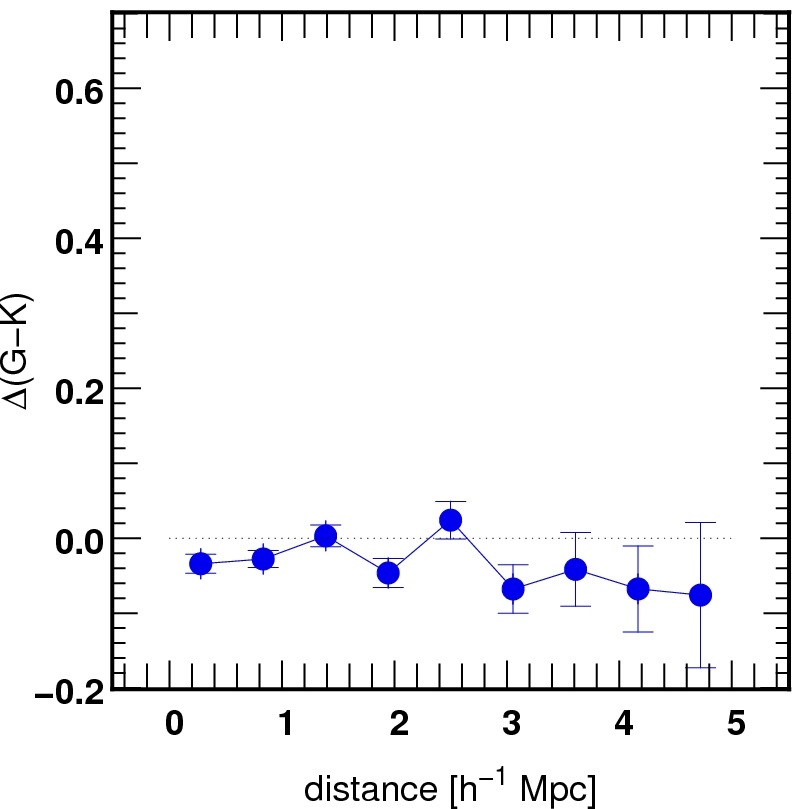}
 \includegraphics[width=0.3\textwidth]{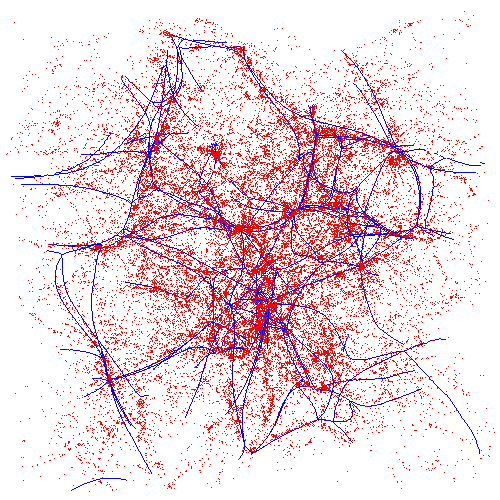} \hfill
  \includegraphics[width=0.3\textwidth]{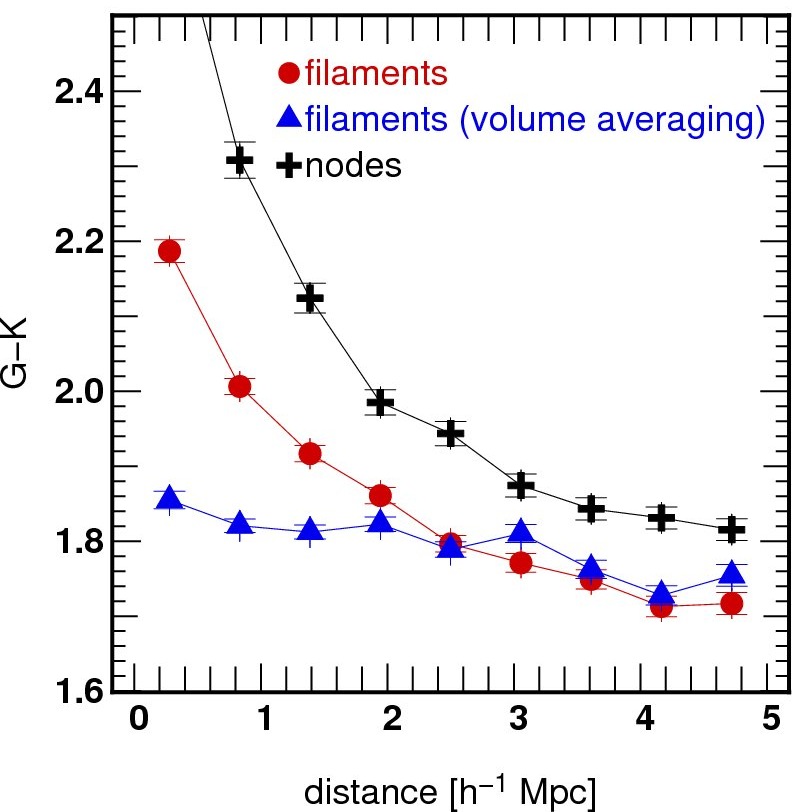} \hfill
  \includegraphics[width=0.3\textwidth]{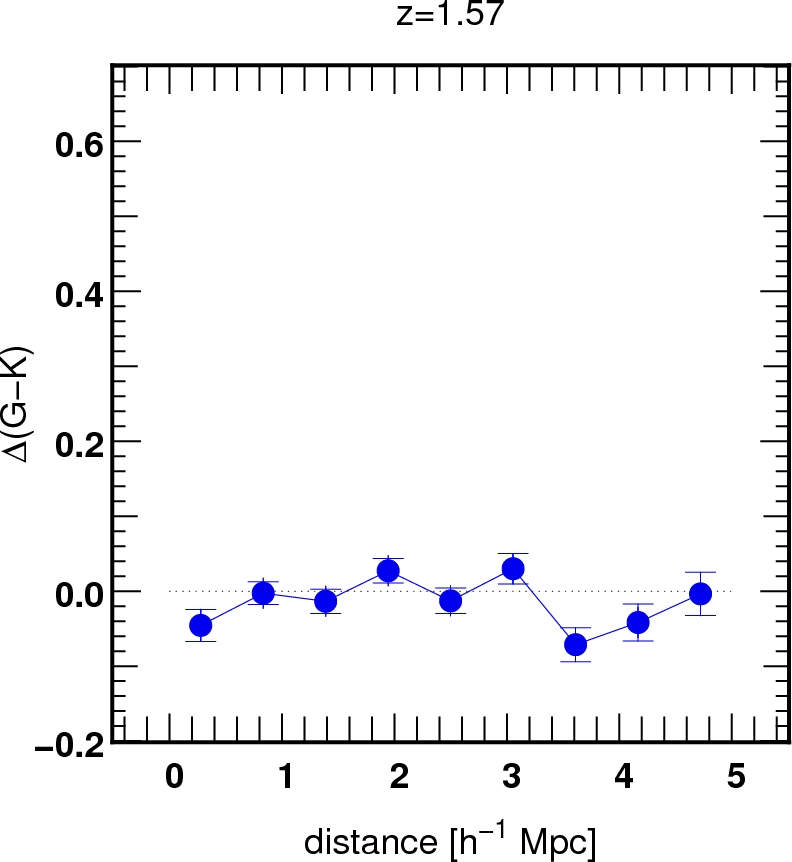} \hfill
\caption{\emph{Left:} Galaxies (\emph{red}) and skeleton (\emph{blue}) for a smoothing length of 8 (\emph{top}) and 16 (\emph{bottom}) pixels. Increasing the smoothing length selects the main features of the cosmic web. \emph{Middle:} Same as Figure \ref{fig:color_all} for a smoothing of 8 (\emph{top}) and 16 (\emph{bottom}) pixels.  \emph{Right:} Same as Figure \ref{fig:color_pairs} for a smoothing of 8 (\emph{top}) and 16 (\emph{bottom}) pixels.}
\label{fig:smooth}
 \end{center}
\end{figure*}

To compute the skeleton, the density field needs to be smoothed to insure sufficient differentiability. The smoothing length is a free parameter, allowing one to probe different scales. All the previous results have been obtained with a Gaussian smoothing over $\sigma=$12 pixels  (with a $256^3$ grid), which corresponds to 2 $h^{-1}$ Mpc. Figure \ref{fig:smooth} shows the results for a smoothing length of 8 and 16 pixels. Increasing the smoothing allows to select only the biggest features of the cosmic web. Thus, it changes the influence of the nodes in two different ways. First it increases the average colour near nodes, the reddest galaxies being located in the biggest clusters. Then it increases the influence scale of the nodes: galaxies properties can be influenced by the main clusters even at distances of several megaparsecs.

Nevertheless, the influence of the filaments is still vanishing when the influence of the clusters is correctly removed. Even the biggest filaments do not have a direct effect on the properties of  galaxies.

\subsection{Spurious filaments}
\label{sec:test_cut}
\begin{figure}
 \begin{center}
 \includegraphics[width=0.85\columnwidth]{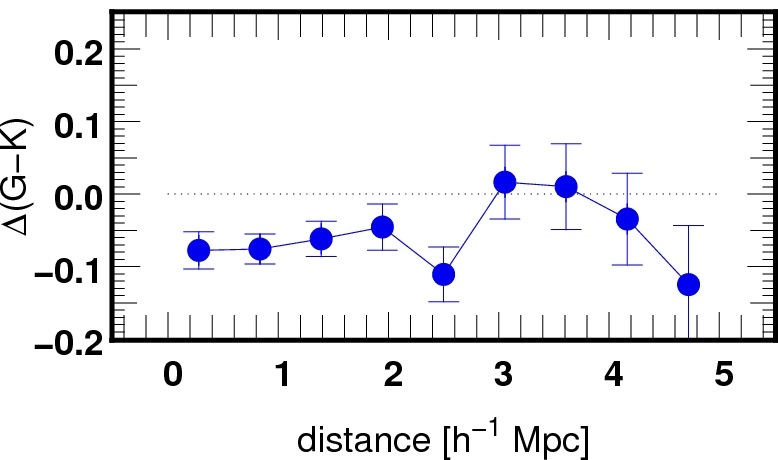}
\caption{Same as Figure~\ref{fig:color_pairs} when the 50$\%$ less-dense filaments are removed. It shows that restraining to the more physical filaments does not change our findings.}
\label{fig:cut}
\end{center}
\end{figure}

The fact that  that some filaments of the skeleton could be spurious and would not correspond to any physical filaments could be responsible for the lack of dependence of spectroscopic properties with the distance to skeleton. This would lead to a dilution of the dependence, since the galaxies near spurious filaments would not be correctly taken into account.
In order to check that the skeleton algorithm is not introducing such spurious filaments, we remove the less physical filaments. Considering whole filaments, \emph{i.e.} a set of contiguous segments between two given nodes, ensure us that this procedure will not depend on the distance to nodes and will not introduce any bias. The selection criterion is based on the density of the underlying field, averaged over the filament. We choose a threshold such that up to 50\% of the skeleton is removed. The result of the pair comparison procedure is given in Figure \ref{fig:cut}. It shows that even when the less significant filaments produced by the skeleton algorithm are removed, the colour of the galaxies does not seem to depend on the distance to filaments. The result presented in the main text is therefore robust and cannot be explained as an artifact of the skeleton algorithm.

\subsection{A toy model for the observed colours}
\label{s:modeldamien}

\begin{figure}
\begin{center}
\includegraphics[width=0.85\columnwidth]{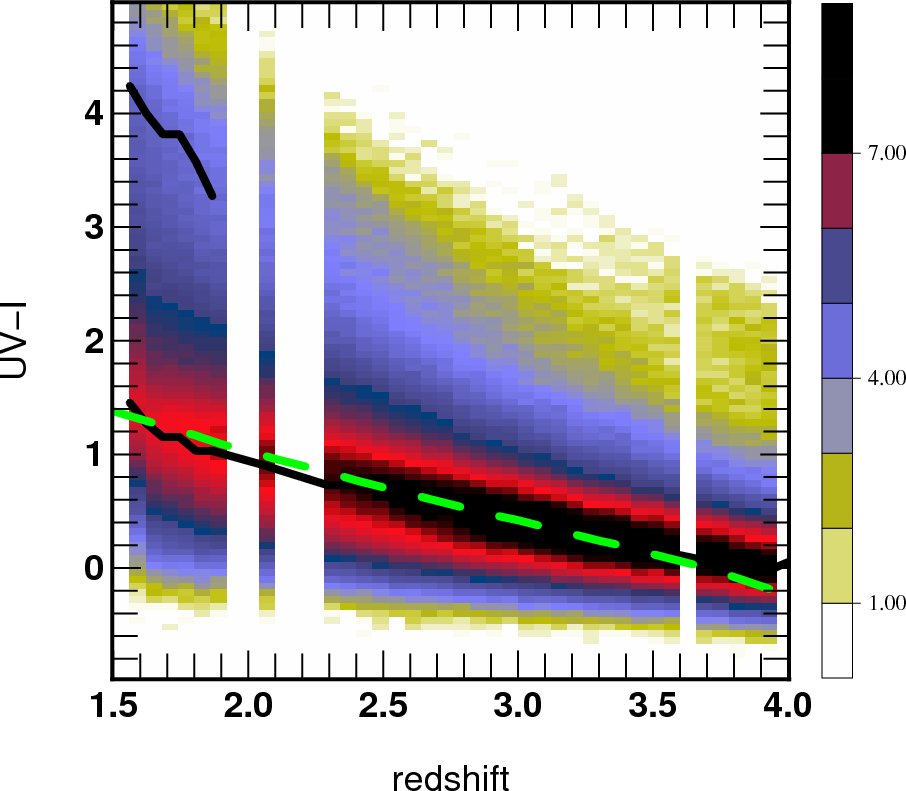}
\caption{Evolution of the distribution of the rest-frame colour. {\sl Black lines}: maximum and secondary maximum of the distribution. {\sl Green line}: {\sc p\'egase} model.}
\label{fig:color_histo_evol}
\end{center}
\end{figure}

Deriving colours for objects in the simulation is a long and complex
process, initiated from a model for the primordial density
fluctuations, and involving huge computing resources to grow the
structures. As mentioned in section \ref{s:MN}, various approximations
and recipes are required to obtain the end-products of interest here,
namely (virtual) galaxies.

To check that the colours presented here make sense with respect to 
what we know of galaxy colours today, we can either compare them to
observations or to previous models. For simplicity, and to avoid
potential biases inherent to observations done at various redshifts, we choose to test
our results against a reasonably simple model of galaxy formation. Another reason to do so is because we did not include dust in the SED modelling of the simulation. A direct comparison with observations is therefore hazardous.

As a basis for this comparison, we use an idealised scenario with a smooth
star-formation history leading to the average colours of local late spiral Sd
galaxies. Such a scenario is presented in \cite{PEG2} or
\cite{PEG-HR}. It involves the infall of
primordial gas from a reservoir onto a potential well at a rate proportional to
$\exp(-t/\tau)/\tau$ with $\tau =\;6$~Gyr. As the accreted gas cools, stars
begin to form with a Schmidt Law, at a rate proportional to the gas
density SFR$= (14 \mathrm{Gyr})^{-1} \rho_{\rm gas}$. Modelled with the code PEGASE.2, the
metallicity of such a galaxy evolves consistently with the yields from
supernovae, and many physical properties are monitored, as well as
predicted spectra and colours. These two time-scales are the only
parameters that were tuned to match the average colours of
local star-forming galaxies with the hypothesis that the redshift of
formation of the first stars is $z=10$. In practice, choosing $z>4$
produces an almost equally satisfying match for the observed colours. 

This scenario, combined with others, is
also successful in reproducing the galaxy counts in optical and NIR
bands \citep{PEG2}. Although such successes are
appealing, they do not prove that the scenario  is correct and it
must be taken with caution.
Still, we venture into the comparison of the MareNostrum simulated colours
with the evolving colours derived from such an idealised scenario. We compare in Figure \ref{fig:color_histo_evol} the evolving
distribution of the UV-I rest-frame colour (which encompasses the 4000
\AA \, break) for galaxies in the simulation with this so-called
``monolithic'' scenario. For consistency with the simulation, we do not include dust in the
model for the Sd spiral.
We find a remarkably good agreement between the colour peak of the blue
sequence and the colour of the scenario at every redshift between
$z=1.6$ and $z=3$, in a range where the refinement of the grid used for the
simulation is comparable from one bound to the other.
It suggests that the bulk of the
galaxies might follow the  path of the idealised scenario and end-up in local late-type spirals.
Such an agreement is comforting and suggests that the colours derived
from the simulation are very reasonable in this redshift range.
We must stress, however, that this comparison does not validate the
monolithic scenario which was mainly tuned to reproduce $z=0$
colours: the match of its colours with distant galaxies is much more
difficult to check because of the difficulty to identify the
progenitors of local spirals at higher redshift.
And of course, this comparison cannot be used to argue that this cosmological
simulation can be reduced to a monolithic collapse, which is contrary
in nature to the hierarchical paradigm.
Nevertheless, we might learn from this exercise that {\it on average}, the
evolution of the population of blue galaxies in MareNostrum is driven by an {\it average} accretion rate following the natural law used above, and
that {\it on average} the Schmidt law described above is
representative of the global star-formation activity.
Still, the exact values of the time-scales used for our scenario
should not be taken for granted: a significant degeneracy exists
between them, not to mention the hypothesis made on the universality
of the IMF.

The main outcome of this test is that the rest-frame colours that we
produce seem coherent, and are consistent, if we extrapolate them to $z=0$, with the
colours of local star-forming galaxies.

\subsection{MareNostrum simulation sanity check}\label{sec:sim-ok}
\begin{figure}
 \begin{center}
 \includegraphics[width=0.95\columnwidth]{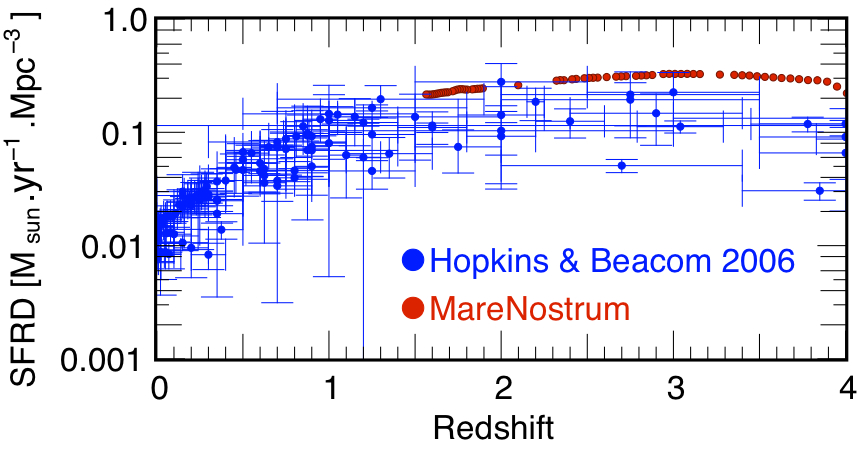}
\caption{Comparison between the star formation rate in the MareNostrum simulation and in observed surveys \citep{Beacom}.
The virtual SFRD seems globally consistent, if somewhat high.
}
\label{fig:sfrd}
\end{center}
\end{figure}

\begin{figure}
 \begin{center}
 \includegraphics[width=0.85\columnwidth]{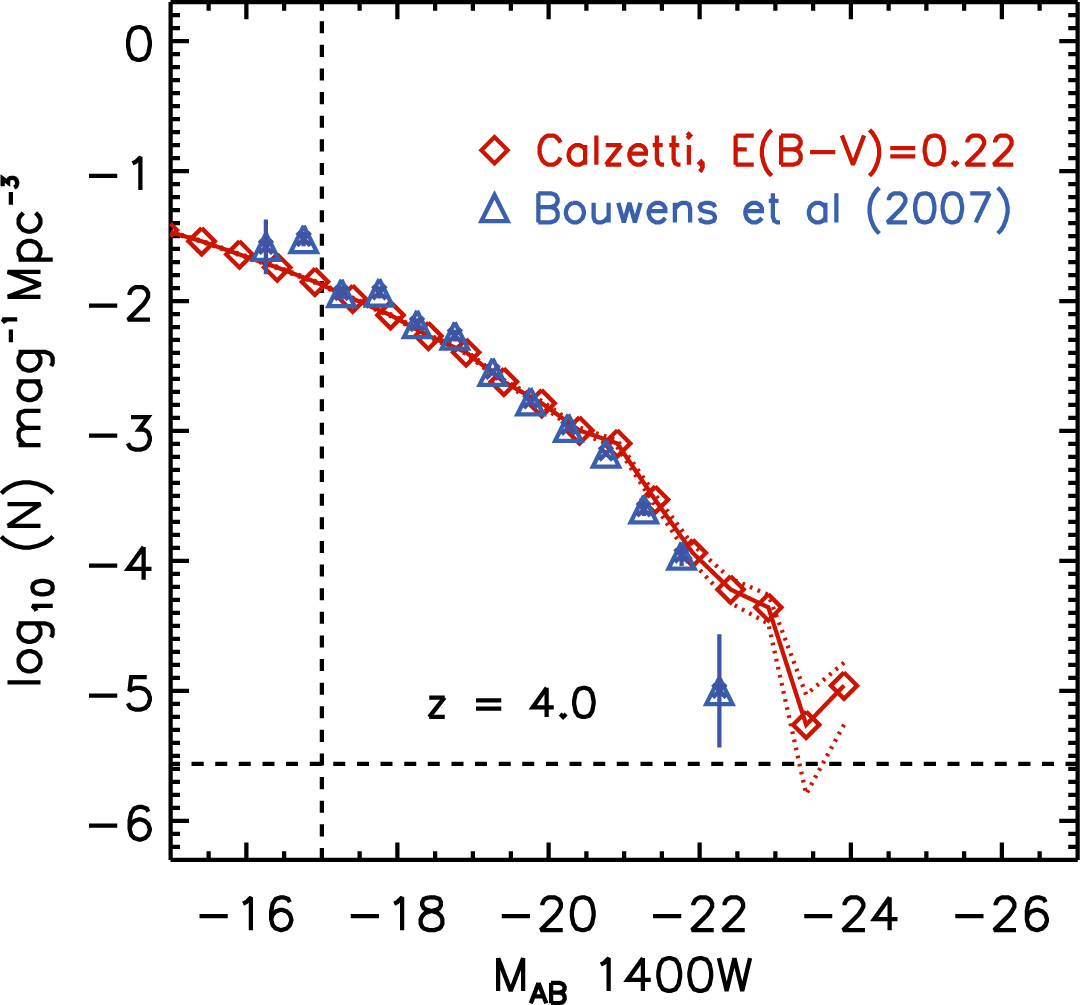}
\caption{ the UV (1400 {\AA}) rest frame
galaxy luminosity function (LF) measured in the Mare Nostrum simulation at redshift 4.
 Red diamond LFs stand for galaxies uniformly
extinguished with a 
\protect\cite{calzetti01}
law. 
Blue triangles correspond to
data gathered by Bouwens and collaborators \citep{bouwens}. 
Vertical dashed lines indicate mass resolution, horizontal dashed lines
volume resolution limits. }
\label{fig:LF}
\end{center}
\end{figure}

Let us carry a couple of checks on the features of the MareNostrum simulation  and its postprocessing using classical probes.
Figure~\ref{fig:sfrd} compares the  star formation rate in the MareNostrum simulation and in the \cite{Beacom} compilation.
The virtual cosmic SFRD seems roughly consistent  with the observed one (which is corrected for dust reddening), although on the high end.
Figure~\ref{fig:LF} compares the UV (1400 {\AA}) rest frame
galaxy luminosity function (LF)  for galaxies  measured in the Mare Nostrum simulation at redshift 4 uniformly
extinguished with a 
\cite{calzetti01} law of $E(B-V)=0.22$,  to that observed by \cite{bouwens}.
Our simulated LFs at z$\sim$4 are in good agreement with the
available UV data  (see \cite{Devriendt2009} for a more detailed comparison).
The apparent discrepancy between the perfect agreement of the UV LFs and the slight overprediction of the SFRD can be explained by several factors: the $z=4$ data points for the SFRD from \citet{Ouchi04} are corrected with a dust extinction measured on individual objects, spanning a range of E(B-V) $\simeq$~0.0 to 0.4. Moreover, other uncertain factors are involved in the comparison such as the correction for the IGM absorption or the model-dependent law to convert from UV luminosity to SFR.

\subsection{Tentative modeling of the effect of dust}\label{sec:dust-ok}

\begin{figure}
 \begin{center}
 \includegraphics[width=0.85\columnwidth]{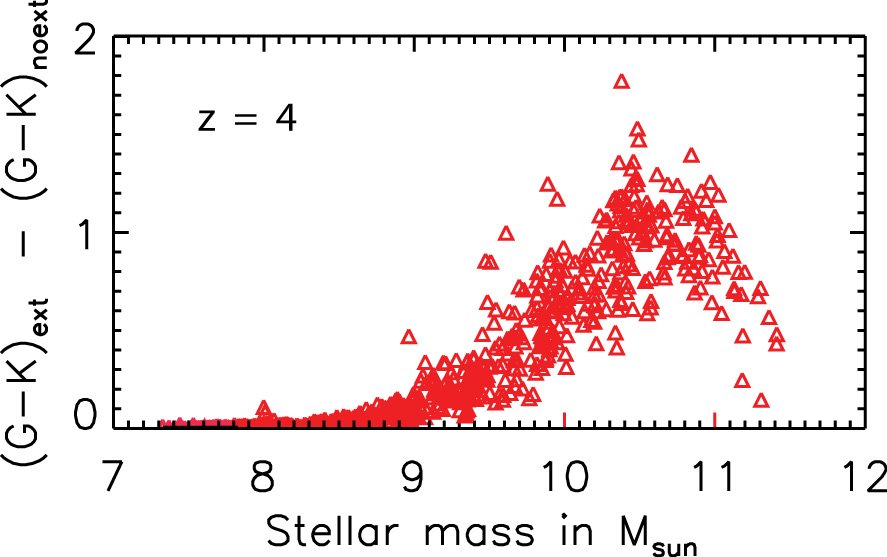}
\caption{ The effect of dust on G-K colours as a function of stellar mass at redshift 4. Dust is modeled from the metallicity of a fair sub-sample of the underlying galaxy following the prescription
of \protect\cite{Devriendt2009}.  Note that the low mass objects responsible for the bimodality described in this paper are not reddened. }
\label{fig:dust}
\end{center}
\end{figure}

\begin{figure}
 \begin{center}
\includegraphics[width=0.85\columnwidth]{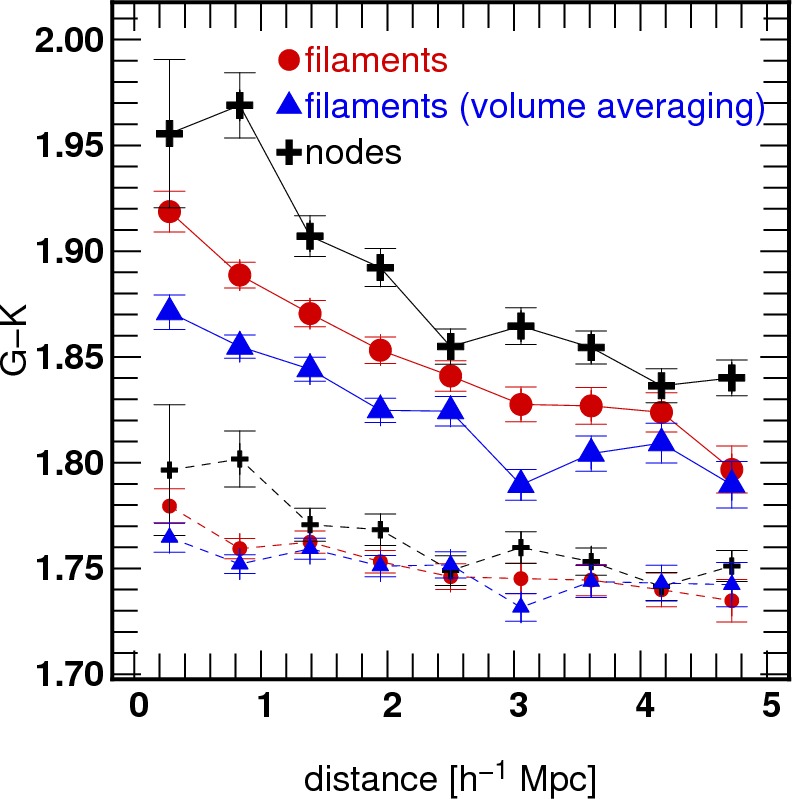}
\includegraphics[width=0.85\columnwidth]{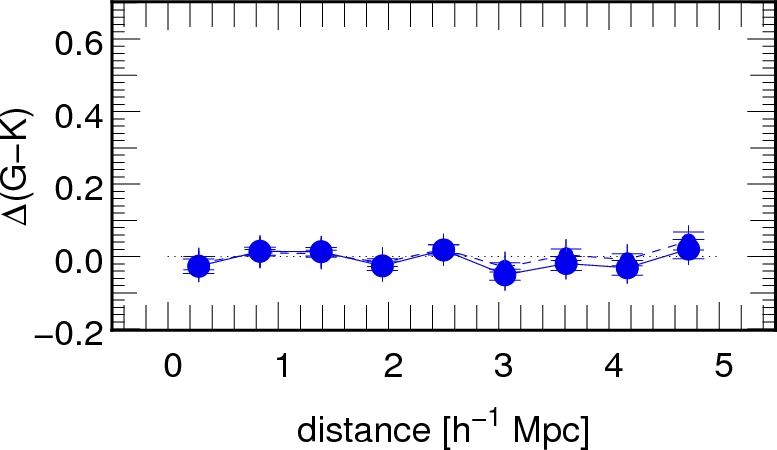}
\caption{Same as Figure~\protect\ref{fig:color_all} (\emph{top}) and Figure~\ref{fig:color_pairs} (\emph{bottom}) at higher redshift (z=4) with ({\sl plain line}) and without ({\sl dashed line}) dust absorpion. The net effect of dust is a global offset of the colours. The colour gradient for the nodes is  slightly stronger with dust, which   reflects the presence of massive dusty galaxies at the nodes, 
but does not modify our conclusions: filaments alone do not have a significant influence on the (dusty) colours of galaxies (\emph{bottom panel}).}
\label{fig:dust-fil}
\end{center}
\end{figure}

Let us finally investigate the effect of dust on the G-K colours computed in this paper. 
Following \cite{Devriendt2009}, we compute the internal absorption of each galaxy using the metallicity of the gas within that galaxy as a proxy.
Figure~\ref{fig:dust} shows the difference in G-K colours between the extinct and the non-extinct galaxies of the MareNostrum simulation at $z=4$ as a function of galactic stellar mass.
It reaches $\sim1$ magnitude at $\sim10^{11} M_\odot$, but remains quite small for the low mass satellites responsible for the observed bimodality.
\\
Figure~\ref{fig:dust-fil}   displays the same colour gradient  as Figure~\ref{fig:color_all} (and  Figure~\ref{fig:color_pairs}) at redshift 4 for {\sl reddened} G-K colours; the corresponding  unobscured colour gradients are shown as dashed lines. Taking into account the dust (i) produces a global shift of the colours, and (ii) increases somewhat the colour gradient  as a function of the distance to nodes, as massive (more obscured, see Figure~\ref{fig:dust}) galaxies are statistically more present at the nodes. Nevertheless the effect of dust on filaments remains much lower than the influence of nodes.

\section{Photometric catalogues}\label{sec:catalogs}

\begin{table*}
\begin{tabular}{l|l|c}
 \textbf{Name} & \textbf{Reference} & \textbf{Mean wavelength}\\\hline
FUV Galex & \cite{Galex} & 1520 \AA\\
G & \cite{SteidelG} & 4810 \AA\\
R & \cite{SteidelG} & 6980 \AA\\
I CFHT 12K & \cite{CFHTI} & 8130 \AA\\
SDSS-z & \cite{SDSSfilters} & 8960 \AA\\
Johnson K & \cite{Johnsonfilters}& 21950 \AA\\
SPITZER IRAC channel 1 IRAC-3.6$\mu$m & \cite{IRAC} & 35610 \AA\\
SPITZER IRAC channel 4 IRAC-8$\mu$m & \cite{IRAC} & 79580 \AA\\
\end{tabular}
\caption{Avalaible filters.}
\label{tab:filtres}
\end{table*}

The data used in this study are made publicly available online: {\tt http://www.iap.fr/users/pichon/MareNostrum/} {\tt catalogues}.
The catalogues contain the properties of the galaxies (position, colour, metallicity, age, SFR, stellar mass) and their environment (distance to skeleton and to nodes, corresponding densities) for redshifts 1.57, 1.8, 2.1, 2.51, 3.01, 3.53 and 3.95. The number of galaxies in each catalogue is respectively 97563, 103589, 111184, 119978, 124642, 119612 and 103187. See Table \ref{tab:catalogue} for a detailed list of the properties. The observed and the rest-frame colours are given by the magnitudes in the 8 filters described in Table \ref{tab:filtres}. Other filters could be implemented upon request.
 \begin{table*}
\begin{tabular}{l|l|c|r|r}
 \textbf{Name} & \textbf{Description} & \textbf{Unit} & \textbf{First} & \textbf{Last}\\\hline
xpos & X position & Mpc/h & 0.201926 & 1.3122\\
ypos & Y position & Mpc/h & 16.6967 & 45.9471\\
zpos & Z position & Mpc/h & 47.9065 & 43.5113\\
redshift & Redshift & - & 1.56506 & 1.56506\\
z\_l & Luminosity-averaged stellar metallicity & dex & -1.69139 & -2.97829\\
z\_m & Mass-averaged stellar metallicity & dex & -1.78234 & -3.03794\\
age\_l & Luminosity-averaged stellar age & Myr & 931.873 & 1429.32\\
age\_m & Mass-averaged stellar age & Myr & 1763.61 & 1640.19\\
FUV & Observed AB magnitude in the FUV band & mag & 28.0128 & 47.3694\\
G & Observed AB magnitude in the G band & mag & 20.9583 & 36.5068\\
R & Observed AB magnitude in the R band & mag & 20.9726 & 35.7524\\
I & Observed AB magnitude in the I band & mag & 20.8269 & 35.2413\\
z & Observed AB magnitude in the z band & mag & 20.6424 & 34.8702\\
K & Observed AB magnitude in the K band & mag & 19.3598 & 33.2805\\
IRAC3p6 & Observed AB magnitude in the IRAC 3.6 $\mu m$ band & mag & 19.1845 & 33.252\\
IRAC8 & Observed AB magnitude in the IRAC 8 $\mu m$ band & mag & 20.0335 & 34.2273\\
FUV\_restframe& Absolute (rest-frame) AB magnitude in the FUV band & mag & -25.3455 & -9.4258\\
G\_restframe & Absolute AB magnitude in the G band & mag & -26.5264 & -12.6057\\
R\_restframe & Absolute AB magnitude in the R band & mag & -23.829 & -12.9374\\
I\_restframe & Absolute AB magnitude in the I band & mag & -26.9103 & -13.0224\\
z\_restframe & Absolute AB magnitude in the z band & mag & -27.0163 & -13.0619\\
K\_restframe & Absolute AB magnitude in the K band & mag & -36.8808 & -12.6582\\
IRAC3p6\_restframe & Absolute AB magnitude in the IRAC 3.6 $\mu m$ band & mag & -26.1501 & -11.8306\\
IRAC8\_restframe & Absolute AB magnitude in the IRAC 8 $\mu m$ band & mag & -24.7171 & -10.2381\\
distance\_skel & Distance to the skeleton & Mpc/h & 0.371031 & 3.73985\\
distance\_node & Distance to the nodes of the skeleton & Mpc/h & 0.811432 & 5.69763\\
sfr & Star Formation Rate & $M_{\sun} / yr$ & 95.8817 & 0\\
stellar\_mass & Total stellar mass & $M_{\sun}$ & 5.11478e+11 & 9.93592e+06
\end{tabular}
\caption{Content of the catalogues. The first and last galaxies of the catalogue at $z=1.57$ are given as reference.}
\label{tab:catalogue}
 \end{table*}
For  the sake of simplicity, the skeleton are saved as a set of coordinates together with the mean dark matter density. 
The node catalogue follows the same prescription.
Here the catalogues are distributed as VOtables\footnote{http://www.ivoa.nest/Documents/latest/VOT.html}, FITS and ASCII.
These could be of use to anyone intending to compare this large-scale simulation to {\sl e.g. } observations at high redshifts.  The skeletons are also available there for further
investigation.

\section{Subgrid physics}\label{sec:recette}
The simulation analyzed in this paper relies on subgrid physics which we now briefly summarize.
\subsection{Star formation rate}
Let us recall the method for implementing star formation in RAMSES as described in \cite{RT}. It is based on a phenomenological approach  adopted in many, if not all, cosmological studies.
It relies on observations of the scaling law between star formation rate and gas densities of local galaxies \citep{Kennicutt1998}.
 Basically, one considers that star formation proceeds at a given time scale, written here $t_*$, in regions where one or several physical criteria are fulfilled. We adopt a simple scheme to turn gas mass into star particles, adding a source term in the continuity equation
  \begin{eqnarray}
    \left( \frac{D\rho}{Dt} \right)_* = & -{  \rho}/{t_*} \,,
    & \mbox{~if~} \rho > \rho_0= 0.1 {\rm cm}^{-3},
    \label{starrateeq} \\
    \nonumber
    \left( \frac{D\rho}{Dt} \right)_* = & 0 \,,& \mbox{~otherwise},
  \end{eqnarray}
 where the star  formation time scale $t_*$ is proportional
  to the local free-fall time
  \begin{equation}
    t_* = t_{\rm 0 }\left( \frac{\rho}{\rho_{\rm 0}} \right)^{-1/2}\,, \quad {\rm with} \quad t_0=2 {\rm Gyr}.
  \end{equation}
This choice of $t_0$ corresponds to the 5 \% efficiency mentioned in the main text.

\subsection{Supernovae feedback}

Following closely \cite{Dubois08}, we define the mass vanished in the star formation process as :
\begin{equation}
\left ( \Delta m_g \right )_{\rm SF}=m_*(1+\eta_{\rm SN} +\eta_W) \, ,
\label{dmgstar}
\end{equation}
where $m_*$ is the final mass of the star particle, $\eta_{\rm SN}$ is the fraction of mass  in supernovae ejecta per solar mass of stars formed ($\eta_{\rm SN}=0.1$ for the assumed Salpeter IMF)
while $\eta_W$ is the mass loading factor that  determines the  gas mass  entrained by the supernovae ejecta (which is set here to 1, see  \cite{Dubois08}). 
We assume that these debris are distributed according to a Sedov blast wave solution with a maximum speed given by
\begin{equation}
u_d={ u_{\rm SN} \over \sqrt {1+ \eta_W/\eta_{\rm SN}}} \, ,
\end{equation}
where
$u_{\rm SN}$ is  the typical velocity corresponding to  the kinetic energy
released in one single  supernova explosion ($u_{\rm SN}\simeq 3200 \, \rm
km.s^{-1}$).
The energy released to the gas by the debris is
\begin{equation}
E_d=\eta_{\rm SN} {m_* \over M_{\rm SN}} E_{\rm SN} \, ,
\end{equation}
where
$M_{\rm SN}$ and $E_{\rm SN}$ are respectively the typical progenitor mass and
energy  of an  exploding type  II  supernova ({\sl i.e.} $M_{\rm SN}=10\,  \rm
M_{\odot}$  and  $E_{\rm SN}=  10^{51}   \,  \rm  erg$).

\label{lastpage}

\end{document}